Mid-Infrared Spectroscopy of Components in Chondrites: Search for Processed Materials in Young Solar Systems and Comets

A.Morlok [a,b,*] (morlokan@uni-muenster.de; Phone: ++49 251 83 39069)

C. Lisse [c] (carey.lisse@jhuapl.edu)

A. B. Mason [d] (andmas@utu.fi)

E. Bullock [e] (bullocke@si.edu)

M.M.Grady [a,f] (m.m.grady@open.ac.uk)

[a] Department of Mineralogy, The Natural History Museum, Cromwell Road, London SW7 5BD, UK

[b] Current address: Institut für Planetologie, WWU Münster, Wilhelm-Klemm-Straße 10, 48149 Münster, Germany

[c] The Johns Hopkins University Applied Physics Laboratory,11100 Johns Hopkins Road, Laurel, MD 20723, USA

[d] Finnish Centre for Astronomy with ESO (FINCA), University of Turku, Tuorla Observatory Väisäläntie 20, FI-21500 PIIKKIÖ, Finland

[e] Smithsonian Institution PO Box 37012, MRC 119 Washington, DC 20013-7012, USA

[f] Department of Physical Sciences, The Open University, Walton Hall, MK7 6AA Milton Keynes, UK

*Corresponding author

**Abstract**

We obtained mid-infrared spectra of chondrules, matrix, CAIs and bulk material from primitive type 1-4 chondrites in order to compare them with the dust material in young, forming solar systems and around comets. Our aim is to investigate whether there are similarities between the first processed materials in our early Solar System and protoplanetary disks currently forming around other stars. Chondrule spectra can be divided into two groups. 1) Chondrules dominated by olivine features at ~11.3 µm and ~10.0 µm. 2) mesostasis rich chondrules that show main features at ~10 µm. Bulk ordinary chondrites show similar features to both groups.

Fine-grained matrix is divided into three groups. 1) phyllosilicate-rich with a main band at ~10µm, 2) olivine-rich with bands at 11.3 µm and ~10 µm, 3) pyroxene–rich with several peaks between 9.3 µm and 11.2 µm. Impact shock processed matrix from Murchison (CM2) shows features from phyllosilicate-rich, amorphous and olivine–rich material. CAIs show melilite/spinel –rich features between 10.2 µm and 12.5 µm.

Astronomical spectra are divided into four groups based on their spectral characteristics – amorphous (group 1), pyroxene-rich (group 2), olivine–rich (group 3) and 'complex' (group 4). Group 2 is similar to enstatite-rich fine grained material like e.g. Kakangari (K3) matrix. Group 3 and 4 can be explained by a combination of varying concentrations of olivine and mesostasis-rich chondrules and fine-grained matrix, but also show very good agreement with shock processed material. Comparison of band ratios confirms the similarity with chondritic material e.g. for HD100546, while the inner disk of HD142527 show no sign of chondrule material.

Comparison between the laboratory infrared-red IR data and astronomical spectra indicate a general similarity between primitive solar system materials and circumstellar dust and comets, especially in the inner disks of young solar systems. However, other amorphous materials like IDP/GEMS have to be taken into account.

## 1. Introduction

The study of material in disks around newly-forming stars can provide a valuable source of information for comparison with the first solids that formed in our own Solar System.

Infrared absorbance studies needed to interpret the astronomical infrared observations have been performed on pure minerals in the laboratory (e.g. Jäger et al. 1998; Rotundi et al., 2002; Koike et al, 2003; Tamani et al., 2006; Morlok et al., 2006a, b; 2008; Bowey et al., 2007; Imai et al., 2009; Pitman et al., 2010).

Here, we focus instead on the components of chondrites – matrix, chondrules and calcium aluminum-rich inclusions (CAIs) – in addition to data from bulk samples of ordinary chondrites. Planetary materials are the 'real thing', showing possible features not to be expected in terrestrial or synthetic samples. Previous analyses of CAIs have been reported by Posch et al. (2007); Morlok et al. (2008), fine-grained chondrite matrix (Osawa et al., 2005; Morlok et al., 2010; Tomioka et al., 2012), and bulk data from chondrites (Sandford and Walker, 1984).

The aim of this work is to compare the mid-infrared spectra of these multi-phase components with astronomical infrared spectra of the dust in circumstellar or protoplanetary disks. The detection of such materials in the ongoing formation of young solar systems would allow a direct link of these observations to the evolution and composition of our own Solar System.

Silicate-rich dust in circumstellar disks has been observed around the solar-mass T Tauri stars (e.g. Olofsson et al., 2009; Watson et al., 2009; Furlan et al., 2011), the heavier Herbig- type stars (2-10 times solar mass MΘ; van Boekel et al., 2005; Juhasz et al., 2010), very low mass brown dwarfs (Furlan et al., 2005; Sicilia-Aguilar et al., 2008) and Be-type giants (15 to 70 MΘ) (Kastner et al., 2010). These astronomical observations provide mineralogical data of dust processing throughout all steps in stellar evolution (e.g. Apai et al., 2010).

In the nearby (up to ~300 pc), young (1-5 Myr) star forming regions like the Taurus-Auriga, Chamaelon, Ophiuchus, Perseus, Lupus and Serpens associations, we see average abundances of crystalline silicates between 16% to 19% for the warm and cold disk parts. This means that pristine amorphous material is still dominant (Olofsson et al., 2010). These amorphous silicates usually have enstatite and forsterite stoichiometry, the crystalline components are mainly enstatite/diopside, forsterite and silica ($SiO_2$). Suggested processes for the formation of crystalline phases are evaporation followed by recondensation in the very inner disk close to the star (< 1AU), annealing by the star or shock processes at larger distances (~10 AU) (e.g. Harker and Desch, 2002; Oloffsson et al., 2009; Watson et al., 2009; Apai et al., 2010).

Higher abundances of crystalline phases in the central parts of the disk indicate that the hot inner region is the main production site (e.g. van Boekel et al., 2004; Schegerer et al., 2009). However, similar crystalline abundances are seen e.g. by Sargent et al. (2009), Olofsson et al. (2010) and Oliveira et al.(2011)in the inner and outer disk, indicating little initial segregation, large scale transportation/ mixing or crystallization in the outer regions.

Transportation of processed crystalline material through radial mixing by turbulent flow has been suggested based on infrared observations of comets (Wooden et al., 1999; 2000; Bockelee-Morvan et al., 2002). Recent solar system comet studies also argue for pronounced mixing in the early disk as comets were being formed (Lisse et al., 2006 , 2007; Brownlee et al., 2006).

Comets are probably the most primitive bodies surviving in our Solar System, thus their mineralogy is also of high interest in the context of this paper (e.g. Wooden, 2008). The GEMS (glass with embedded metal and sulfides) dominating the amorphous silicates in anhydrous interplanetary dust particles (IDP) of cometary origin show similarity to pristine materials (e.g. Alexander et al., 2007). So cometary dust is probably more pristine than primitive meteorite materials. However, chemical and isotopic data of most GEMS point towards a formation as condensates in the early Solar System (Keller and Messenger, 2011). In any case, similar materials can also be expected in other young, forming solar systems.

Studies on IDP were made e.g. by Sandford and Walker (1985) or Brunetto et al. (2011), and on cometary material directly collected via the Stardust probe, e.g. by Rotundi et al. (2008).

Some care has to be taken however, when comparing laboratory spectra with astronomical observations. What do we actually observe in the astronomical spectra?

Except for the studies by van Boekel et al.(2004) and Okamoto et al.(2004) which directly probed the inner and outer disk of three Herbig stars and debris disk Beta Pictoris, available mid-infrared spectra are basically of point sources without direct spatial information. Still, some information about the location of the dust can be extracted. In T Tauri disks, the 10 µm range of the spectra comes from the hot, inner part (<3 AU). The ~20 µm part originates in the upper disk layer of the comparatively colder, outer regions (up to ~25 AU). In the heavier Herbig stars with bigger disks the 10 µm zone ranges from 0.5-50 AU, in the low–mass brown dwarfs at <0.1 AU (e.g. van Boekel et al., 2005; D'Allessio, 2006; Kessler-Sillacci et al., 2007; Oloffsson et al., 2009).

A further limitation is the emission efficiency. The spectral features decrease with increasing grain size. The signal in the 10 µm region of an infrared spectrum comes mainly from grains smaller than ~2 µm, while larger grains up to 6 µm provide the features at higher wavelengths up to 40µm (Min and Flynn, 2010). Thus any information regarding processing of larger grains could be lost, although material from larger grains could be recycled by fragmentation (e.g. Olofsson et al., 2009).

In addition, the infrared radiation we observe originates in the hot surfaces of the disks. The colder mid-plane, where most of the mass resides, is only accessible at longer wavelengths (mm range), where structural information of the dust is difficult to obtain (Min and Flynn, 2010). Chondrules, CAI, fine-grained matrix and other meteoritic components probably form in the mid-plane. The accretion of larger bodies and finally planets also takes place here (Apai and Lauretta, 2010). Recent studies explain the existence of micron sized dust over the long observed time range for circumstellar disks with a replenishment of the dust by vertical mixing from the mid-plane and fragmentation by the grains in collisions (e.g. Dullemond and Dominik, 2005; Kessler-Silacci et al., 2006; Olofsson et al., 2009).

This processing could also produce micron-sized dust from chondrules and CAI, which in their final form would be too big for characteristic emission features in the mid-infrared. Such fragmentation of processed components is also supported by meteorite studies (e.g. Jones, 1996; Scott and Krot, 2005a and b; Ruzicka et al., 2008).

The aim of our study is to compare chondrule, CAI and fine-grained matrix mid-infrared laboratory absorbance spectra to data from nearby forming solar systems, and to find if there are any similarities between the two groups. Together with earlier studies focusing on single chondrite components such as CAIs (Morlok et al., 2008) and experimentally shocked/heated matrix (Morlok et al., 2010; Tomioka et al., 2012), this paper aims to provide a series of spectra for this kind of studies.

**2. Materials and Methods**

2.1 Sample Selection

The chondrules presented in this study represent the most common types found in chondrites (Lauretta et al., 2006) (Fig. 1a-c). Chondrules were separated from the fine-grained matrix of an Allende bulk sample (sample number Natural History Museum, London BM1969, 148) (Fig.1). Allende (CV3.2) was collected directly after its fall, and as a result, has undergone limited terrestrial alteration. Also, large quantities of sample material are easily available. Chondrites of type 3 were only slightly affected by parent body processes such as aqueous alteration and thermal metamorphism, so the components (chondrules, CAI and matrix) are basically pristine – they have not undergone extensive changes in their structure and composition since their formation.

We also included bulk analyses of the ordinary chondrites Parnallee LL3.6 (BM34729), Ceniceros H3.7 (BM1990.M6) and Baratta L3.8 (BM1927, 1287). The ordinary chondrites are very common, making up 85% of all falls (Weisberg et al., 2006). The samples used in this study cover the range of the ordinary chondrites from low iron/metal contents (type LL) to high abundances (H). Chondrule abundance in this group is very high with 60-80 volume%, while fine-grained matrix (<15 volume %) and CAIs (usually <<10 volume %) are only minor components (Brearley and Jones, 1998). It is difficult to physically separate the various components in ordinary chondrites, so we use the bulk spectra as representative for this group. The small pyroxene-rich enstatite chondrite group is represented by Indarch EH4 (BM1980.M2). The type 4 designation indicates thermal metamorphism to a high enough temperature that the amorphous chondrule mesostasis began to devitrify and the chemistry equilibrates due to diffusion (Brearley and Jones, 1998).

We also obtained spectra of fine-grained matrix material to fill gaps in the database. The matrix material was separated from 'bulk' carbonaceous chondrites; we avoided fine-grained rims close to chondrules. For CM2 chondrite Cold Bokkeveld (BM1727), CR2 chondrite Al Rais (BM1971, 289) and CI1 chondrite Alais (BM61329), the decreasing type number indicates the increasing impact of aqueous alteration (Fig.1d-f). On the parent body in the early Solar System, circulation of fluids oxidized the pristine minerals into hydrated phases such as phyllosilicates, iron oxides or carbonates. Type 1 chondrites such as Alais consist entirely of such secondary phases, while type 2 chondrites still contain anhydrous material. Matrix from K3 chondrite Kakangari (BM69062) was also analyzed since it is has not been affected by thermal or aqueous processing. The matrix-rich (up to 77 volume %) Kakangari grouplet has petrological characteristics similar to carbonaceous and ordinary chondrites, but its isotopic characteristics identify the meteorites as separate group (Weisberg et al., 1996).

In order to provide a more comprehensive study, we also include data from earlier studies. This includes spectra of fine-grained matrix from type 3 chondrites Allende CV3.2, Vigarano CV3.3, Ornans CO3.3, Ningqiang and DHO015 CK3, Gjuba CB3 and Acfer207 CH3 (Osawa et al., 2005; Tomioka et al., 2012 and Morlok et al., 2008) (Fig.1g-i). The data from Osawa et al. (2005) was also obtained with a comparable FTIR microscope technique, but at a wider wavelength range (2 to 25 μm).

Analyses of CAIs from Morlok et al. (2008) are all from type 3 carbonaceous chondrites collected directly after they fell (Allende CV3.2, Vigarano CV3.3, and Ornans CO3.3) (Fig.1j-l), and so have undergone minimal terrestrial alteration. Like chondrules, CAIs show a wide range of types and composition (e.g. Brearley and Jones, 1998). While the spectra presented here are representative for typical CAIs, it is not possible to cover the entire range of CAI in the scope of this study. However, CAIs are usually only a minor component in chondrites. These spectra were obtained with the same techniques as the fine-grained matrix material in this study. The spectra are from specific parts of each CAI: 'RIM' indicates material from the outer portions of the CAI, while 'Interior' (INT) is material from the 'core' of the inclusion.

A major process to produce dust in a young Solar System is the collision of planetesimals. Thus the potential effects of the collision on the dust have to be taken into account. This motivated the inclusion of experimentally shocked fine-grained matrix material from the Murchison CM2 chondrite (Morlok et al., 2010). In laboratory shock experiments, material was exposed to shock pressures of up to 49 GPa (Tomeoka et al., 1999). The impact shock transformed the starting material (mainly serpentine) into an amorphous silicate, followed by melting and recrystallization above 36 GPa. The analyses were obtained with a FTIR microscope technique similar to that in this study, in a wavelength range from 2 to 23 μm and a resolution of 4 $cm^{-1}$. For more details see Tomioka et al. (2007) and Morlok et al. (2010).

### 2.2 SEM Analyses of Matrix and Chondrules

In order to characterize the chondrules and matrix, we prepared polished blocks for the analysis with a Scanning Electron Microscope (SEM). Samples were mapped in backscattered electron (BSE) mode, in order to determine the structure of the chondrules. For identification of dominant minerals based on stoichiometry, chemical compositions were measured using energy dispersive X-ray spectroscopy (EDX). A Jeol 5900LV ProbeSEM with a probe current of 2 nA at20 keV and an INCA EDX system at the EMMA labs (Natural History Museum, London) was used. For calibration, both natural and synthetic standards were used. Detection limits of the analysis are at about 0.02 wt.%. The results are shown in table 1a-c.

Chondrules were classified based on their texture and chemical composition. Porphyritic chondrules show large crystals embedded in the amorphous mesostasis. Chondrules of this type that are dominated by olivine are classified as porphyritic olivine (PO), those with mostly pyroxene as porphyritic pyroxene (PP). Intermediate cases are porphyritic olivine and pyroxene (POP). Special cases showing elongated olivine crystals in the mesostasis are barred olivine (BO) chondrules. Fine needle-like pyroxenes radiating from specific points in the chondrules are classified as radial pyroxene (RP) chondrules (Brearley and Jones, 1998).

For chemical classification, the iron and magnesium content in the olivines and pyroxenes was used. Chondrules with forsterite or enstatite contents of over 90% were classified as type I (iron poor), the remaining chondrules as type II (iron rich) (Brearley and Jones, 1998; Lauretta et al., 2006).

### 2.3 FTIR Analyses

The infrared spectra in the observed dust originates from grains with a size <10 μm. So the material has to be ground to a comparable grain size of ~1 μm for laboratory studies. Variations in grain size affect the shape and relative intensities of the infrared bands. These are decreasing band intensity with increasing grain size (Hofmeister et al., 2000; Min et al., 2005).

Spectra for this study were taken in the EMMA labs at the Natural History Museum in London. For bulk transmission analyses we used a Perkin Elmer Spectrum One IR workbench. The wavelength range measured was 2.5–25 µm, with a spectral resolution of 1 $cm^{-1}$. We ground ~2 mg of sample material in an agate mortar for 60 minutes, and mixed the rock powder with ~100 mg KBr, following a well-tested preparation technique (Chihara et al., 2002; Koike et al., 2003; Imai et al., 2009). The mixture was compressed at 10 tons/$cm^2$ for 30 seconds in an evacuated steel pellet die to 1 cm diameter pellets. Differences in band positions and shape occur when different preparation methods are used to grind bulk materials. They are minor and provide comparable spectra in the mid-infrared for hand-grinding such as used in this study, and longer grinding techniques using ball-mills. However, at higher wavelengths (>15 µm) differences are stronger. A discussion of the effects of preparation on shape, porosity and grain size is found in Imai et al., 2009 and Lindsay et al., 2013.

For FTIR microscope measurements of fine-grained matrix, we separated material from polished samples under a binocular microscope using a fine tungsten needle. The sampling areas represent parts clear of visible larger components in the size of several $mm^2$. A Perkin Elmer AutoImage microscope was used for analyses in the 2.5–16.7 µm range, at a spectral resolution of 4 $cm^{-1}$.

For the microscope analyses, the sample material was crushed into a powder with a grain size of ~1µm. This was done using a commercially available diamond compression cell (DCC). The empty diamond the cell was used for background measurements. This technique provides results comparable to the analysis of powders dispersed in KBr (Hofmeister et al., 1997; Osawa et al., 2001; Morlok et al., 2006b). However, the two different types of analyses can show minor differences in, for example, relative band intensities (Hofmeister et al., 2000; Pitman et al., 2010). A potential pitfall of this technique is that there is no exact control of the thickness of the compressed layer, and thus the grain size of the analyzed particles. However, the feature intensity of the raw FTIR spectra is always clearly below an absorbance of 1. This indicates that the sample layer was thin enough to avoid overloading of the spectrum, which prevents major artifacts (Hofmeister et al., 2000). Also, the grain size of the fine-grained matrix material used usually ranges from ~1 µm into the nanometer scale, with only few larger grains (Pontoppidan and Brearley, 2010).

The results are presented in absorbance units for better comparison with astronomical data in Table 2 and Figure 2.

In order to make comparisons with astronomical spectra easier, we averaged the single spectra of several groups of materials with similar bands. In this procedure, we normalized each single spectrum to the strongest feature in the 8 to 13 µm region, and obtained a straight point-by-point mean of all the individual spectra. So each spectrum has the same weight in the averaged spectrum.

The high number of features in most spectra made spectral decomposition difficult. Band positions were obtained in two ways. The first was to simply measure the actual position of a peak in the spectrum without any fitting, by 'eye', using Origin Pro 8. Uncertainties of the band positions are at 0.05 µm (Morlok et al., 2006b). To obtain further characteristics of the features we fitted Lorentzian functions to the bands using the Origin Pro 8 software package. However, using Lorentzian functions for the fitting assumes symmetric shapes of a feature, which may not always be the case. Also, potential unresolved bands or shoulders may affect the decomposition. Very weak bands were impossible to resolve with this technique. Therefore, we present both the 'normal' and fitted, 'barycentric' band positions in addition to the FWHM (Full Width at Half Maximum) and the intensity, which is based on the integrated area of a feature (Pitman et al., 2010). In the comparisons with the astronomical spectra, we always use the 'normal' band position, since this set of data includes also the weak bands and shoulders. In the following comparisons between laboratory spectra and astronomical data (4.1. Grouping of Astronomical Spectra), we regard band positions as similar in the 8-13 mm range when they fall into a range of 0.2 µm. This allows for smaller variations due to slight differences in composition or analytical accuracy.

**2.4 Astronomical Spectra**

Astronomical spectra were selected with the aim of using a diverse group of spectra of circumstellar disks and comets for the comparison with the laboratory data. A further criterion was a high signal to noise ratio.

The astronomical spectra for HD142527, HD163296 and HD144432 were obtained using the mid-Infrared Interferometric Instrument (MIDI) installed at the Very Large Telescope Interferometer (VLTI) at the European Southern Observatory (ESO) in Chile (Van Boekel et al., 2004). Spectra for GK Tau, V955 Tau, DK Tau (Furlan et al., 2011), Oph3/RXJ1612.6-1859 (Bouwman et al., 2008), EK Cha/RECX-5 (Sicilia-Aguilar et al., 2009), and cometary spectra 73P/Schwassmann- Wachmann (Sitko et al., 2011), 9P/Tempel1(post-Deep Impact ejecta) (Lisse et al., 2006 and 2007) and 17P/Holmes (Reach et al., 2010) were obtained using the *Spitzer Space Telescope*. Data for HD100546, HD179218, HD104237 and comet Hale-Bopp are from the SWS of the *Infrared Space Observatory* (ISO) (Sloan et al., 2003; Crovisier et al., 1997). The 8-13 μm spectra for HD179218 and HD144688 were made using the TIMMI2 instrument at the 3.6 m telescope of ESO, La Silla, Chile (van Boekel et al., 2005). Astronomical spectra in figure 3 and 4 are in flux (Jy), compared to the laboratory spectra in absorbance units. To prevent information loss, we avoided smoothing and normalization of the data from laboratory studies and various astronomical sources to the same resolution. For comparison purposes, the spectra were scaled in the 8-13 μm regions to the same intensity. The normalization procedure for figure 5 is explained below. Generally, the data from astronomical sources has a lower signal to noise ratio compared to laboratory data. Also spectra processing of astronomical data can lead to artifacts e.g. by joining data from different spectral ranges.

In order to quantify the similarities between laboratory spectra and astronomical data, we compared the ratios of the peaks at ~10 μm and ~11 μm for laboratory data and selected astronomical spectra. These two features are important in the classification of the astronomical spectra and planetary materials (see 4.1 Grouping of Astronomical Spectra). To allow a direct comparison, both data sets had to be normalized.

We removed the continuum in the 8-13 μm region by dividing the flux ($F_\nu$) by a blackbody continuum ($B_\nu(T)$) calculated using a Planck function for the dust temperature T and radius r of the inner disk for of HD142527 (T=523 K, r=0.3 AU), HD144668 (517 K, 0.5 AU) and EK Cha (T=362 K, 0.1AU (van Boekel et al., 2005; Kessler-Silacci et al., 2007, Sicilia-Aguilera et al.,2009 , Verhoeff et al., 2011). For similar processing of HD100546 and comets Hale Bopp, post-Deep Impact ejecta of 9P/Tempel1 and 17P/Holmes see Lisse et al. (2007) and Reach et al. (2010), for 73P/Schwassmann-Wachmann Sitko et al. (2011).

For a direct comparison of the resulting emissivity data with the absorbance laboratory spectra (Fig.5),

we also base lined the spectra in 8-13 µm region. We used a simple linear base line, between the low points of the silicate intensity in the 6 µm to 13 µm range. After this, all spectra were normalized to 1 using the strongest feature in the 8-13 µm region. Since this comparison is mainly about the relative intensities of the silicate features, we regard this simple method as appropriate (Fig.5a and b).

## 3. FTIR Results

### 3.1 Mid-infrared features of individual minerals

In olivine, the strongest peak in the 8-13 µm region shifts with increasing iron content from 11.2 µm in the magnesium-rich end member forsterite to 11.4 µm for the iron-rich end member fayalite. In a similar way, another strong band shifts from 10.06-10.57 µm. There are two weaker features at 10.4-10.9 µm and 11.9-12.1 µm. At higher wavelengths, strong peaks are at 16.4-17.7 µm, 19.7-20.9 µm, 23.7–27.7 µm, 27.5–31.9 µm and 33.8–38.9 µm, all shifting with increasing iron content (e.g. Koike et al., 2003; Morlok et al., 2006a).

Magnesium-rich pyroxene (enstatite) shows several prominent bands in the 8-13 µm range. The peak shifting with increasing iron content from 9.2 µm to 9.5 µm is especially useful, since it does not overlap with olivine bands in mineral mixtures. Two further strong features shift from 10.7-10.5 µm and 11.7-11.4 µm. Minor peaks in this range occur at 8.7 µm, 9.9-10.4 µm, 10.3 µm and 11.1-11.3 µm. Many weaker bands fall into the range from 12.5 µm to 16 µm. Prominent peaks at higher wavelengths are at 19.5-20.4 µm and 28.1-31.7 µm (e.g. Chihara et al., 2002; Bowey et al., 2007). Calcium-rich pyroxenes (diopside and augite) show similar bands. Most prominent in the 8-13 µm range are features at 9.3 µm, 10.3 µm and 11.4 µm. Further strong peaks at higher wavelengths are at 19.3.-19.5 µm, 20.6 µm, 24.3-25.1 µm, ~30 µm and 32-34 µm (Koike et al., 2000).

Serpentine shows features at 6.0-6.3 µm (water), 9.8-10.2 µm and 15.4-16.2 µm (Morlok et al., 2010). The CAI phase gehlenite (Al-rich melilite) has prominent bands at ~10.3 µm, 10.9–11.2 µm and at 12.3−12.5 µm, plus minor features at 9.4 µm, 9.8 µm, 11.6 µm, 14.0 µm and 15.3µm (Chihara et al., 2007). Spinel, another typical CAI phase, shows a characteristic band between 14.0 µm and 14.9 µm, dependent on the iron content (Chihara et al., 2000, Morlok et al., 2008).

Amorphous silicates show simple spectra in the mid-infrared. A prominent feature falls between 9.0 µm for the $SiO_2$ rich end-members silica and obsidian to 10.8 µm for melilite type glass. Glass with an olivine or pyroxene composition has its strongest band between 9.9 and 10.2 µm. A further major peak is in the 18-20 µm regions (Speck et al., 2011). Polycyclic aromatic hydrocarbons (PAH), organic compounds found in some carbonaceous chondrites, have features at 6.2, 7.7, 8.7 and 11.3 µm (Lisse et al., 2007).

### 3.2. Chondrules

Based on the composition and texture of their major ferromagnesian silicates, the chondrules in our study can be classified as porphyritic olivine (PO; chondrule 1, 2, 4, 9 and 10), barred olivine (BO, chondrule 7), porphyritic olivine and pyroxene type (POP; chondrule3, 5, 11), and radial-pyroxene (RP, chondrule 8) or porphyritic pyroxene (PP; chondrule 6). With the exception of chondrules 8 and 9, all chondrules in this work are FeO-poor type I (Lauretta et al., 2006) (Table 1a, Fig.1a-c).The infrared data confirms this (Table 2, Fig.2). Most spectra show the dominant 11.2-11.3µm and 10-10.2 olivine features as the strongest bands, with secondary peaks at 16.4-16.6 µm, 19.6-19.8 µm, ~24µm and ~34 µm (Koike et al., 2003; Morlok et al., 2006a). Signs of increased pyroxene content in the spectra of the POP group are the emerging crystalline pyroxene shoulders and bands at 9.3-9.4 µm (Bowey et al., 2007). Chondrule 5, which also belongs in this category, only shows a weak pyroxene feature. The classification using SEM/EDX is based on a 'slice' of the chondrules, while the infrared analyses are bulk analyses. Thus differences in the deeper part of the sample become clear in the FTIR analyses, which represent the whole chondrule.

The infrared spectra of chondrules 6, 8 and 9 are characterized in the 8-13 µm regions by a dominating, strongest feature at 10-10.2 µm, with weaker bands at higher wavelengths. This is an indication of a high content of amorphous material (Morlok et al., 2010; Speck et al., 2011), in this case mesostasic phases. The other typical phases in these samples are olivine with weak bands at ~11.2 µm (chondrule 6, 8, 9) and minor pyroxene with features between 12.5 and 16 µm in chondrule 6 (Chihara et al., 2002).

### 3.3 Bulk Chondrites

The spectra of our type 3 ordinary chondrites (Parnallee LL3.6, Ceniceros H3.7, Baratta L3.8) are dominated by strong bands at 10.2 µm, 11.3 µm, 16.9 µm, 20 µm, 24.4-24.5 µm and 28.4-28.7 µm. These are identified as forsterite features (e.g. Koike et al., 2003).

The only clear pyroxene band is a peak at 9.5 µm (Chihara et al., 2002) (Tab.2, Fig.2). The enstatite chondrite Indarch (EH4) shows more features, including typical enstatite bands. The strongest characteristic bands are a double peak at ~9.3 µm, and further bands at 9.8 µm, 10.7 µm, 11.7 µm (Chihara et al., 2002; Bowey et al., 2007) (Tab.2, Fig.2). The pronounced slope of the silicate feature shows a decrease in intensity from the 9.3 µm peak towards longer wavelengths, indicating an additional phase with few distinct bands in the mid infrared, e.g. the Fe metal found in EH chondrites (Brearley and Jones, 1998).

### 3.4. Fine-grained Matrix

EDX data for the composition of the matrix material obtained in this work is very similar to results from earlier studies (McSween, 1977; Rubin et al., 1988; Rubin et al., 1988; Zolensky et al., 1993) (Tab.1b). Only Kakangari (K3) matrix shows variations compared to the literature data (McSween and Richardson, 1977). This is not surprising given the heterogeneity of the fine-grained matrix in this meteorite (Fig. 1g-i). The separated matrix analyzed in this work covers material from two different lithologies, and should be representative of the bulk matrix. For detailed information about DHO015 (CK3), Gujba (CB3) and Acfer 207 (CH3) see Osawa et al. (2005).

The mid-infrared spectra of fine-grained matrix can be divided into three groups (Fig.1d-i).

1) Aqueously altered type 1 and 2 materials (e.g. Alais CI1, Cold Bokkeveld CM2 and Al Rais CR2). These show very similar spectra: strong bands at 9.8-10 µm and 15.2-15.8 µm, mainly serpentine with possible tochilinite features. Weaker peaks between 6.1 and 6.9 µm are due to water and carbonate contents (Osawa et al., 2005; Morlok et al., 2010) (Figure 2). No features from minor phases such as magnetite or sulphides are visible in this part of the spectrum.

2) Pristine material from type 3 chondrites (Fig.1g-i) Ningqiang, DHO015 (both CK3), Ornans (CO3.3), and two Vigarano samples (CV3.3). This material is dominated by strong olivine bands at 10.1-10.5 µm and 11.2-11.4 µm, and a weaker feature at 12 µm (Tab.2) (Koike et al., 2003; Morlok et al., 2006). Ningqiang and Vigarano show pyroxene bands at 9.3 and ~14 µm (Chihara et al., 2002).

Features of further components such as metal and sulphides do not show bands in this spectral region and would mainly add to the featureless continuum. Some weak peaks between 6.2 and 6.9 μm in Ornans are water and carbonates, possibly terrestrial alteration (Osawa et al., 2005; Morlok et al., 2010).

3) The third group includes samples from CB3 chondrite Gujba and CH3 chondrite Acfer 207 (Osawa et al., 2005). They show a mixture of olivine and pyroxene bands at 9.34, 10.6-10.7 and 11.2 μm (Chihara et al., 2002; Morlok et al., 2006a). The spectrum of matrix from K3 chondrite Kakangari has characteristic features at 9.29, 9.88, 10.59 and 11.11 μm, as well as weaker bands at 13.7, 15.5 μm. These are typical bands of iron poor enstatite, the dominant phase of K type chondrites. The second important phase, olivine, is indicated by a peak at 11.1 μm (Brearley, 1989; Chihara et al., 2002; Morlok et al., 2006a).

### 3.4 Shocked Matrix

Infrared features of experimentally shocked matrix material from Murchison (CM2) can be divided into four groups (Tomeoka et al., 1999; Morlok et al., 2010). Group 1, 'Unshocked' material shows similar features to the fine grained matrix from the other type 2 chondrites (Tab.2, Figure 2).
Group 2 (Medium Shocked) is a mixture of amorphous and phyllosilicate material with additional minor olivine that underwent pressures between 21-34 GPa. It shows features of amorphous silicate and serpentine at 9.8 μm, 11.15 μm, between 14.9 and 16.2 μm (Morlok et al., 2010; Speck et al., 2011). Stage 3 (High Shock; 36 GPa) is dominated by Si glass and olivine, with characteristic bands at 11.25 μm, 10.7 μm and 19.5 μm (Morlok et al., 2010). Re-crystallized material (Maximum shocked) at 49 GPa ('Maximum Shock') shows olivine bands at 10.17 μm, 11.27 μm, 16.9 μm and 19.6 μm (Koike et al., 2003).

### 3.5. Calcium-Aluminum Rich Inclusions (CAIs)

The selection of pristine, unaltered CAIs from Ornans CO3.3 and Vigarano CV3.3 (Morlok et al.,2008) shows strong bands at 9.29, 10.25 and 11.26 μm characteristic of Ca-rich pyroxene material e.g. from the outer rim of the Ornans CAI (Koike et al., 2000) (Tab.2, Fig.1j-l).

Melilite-rich material is also found in CAIs (e.g. Ornans CAI Interior, Vigarano CAI1 and 2). This phase shows a variety of strong bands e.g. at 10.3 µm, 10.9-11.3 µm, and 12.3-12.5 µm (Chihara et al., 2007; Morlok et al., 2008).

Spinel, another typical component of CAIs, is characterized by a broad feature that peaks between 13 and 15 µm in Ornans CAI Interior and Vigarano CAI1 and 2. The exact position depends on the Fe content of the material (Chihara et al., 2000; Morlok et al., 2008).

## 4. Discussion

We identify types of astronomical spectra based on the prominence of the observed (mainly ferromagnesian silicate) emission features. The various types of spectra occurring in the circumstellar disks, transitional disks, and comets can be sorted into groups based on their features. This could indicate a general similarity in the formative processes that occurred (see below for statistics). In order to make the discussion simpler, we group spectra with similar features (e.g. a clear peak or band, in contrast to shoulders) in the 10 µm region together.

These groups do not imply evolutionary sequences or genetic relationships; they are used entirely for presentation purposes, to avoid redundancy. Also, the mineral modal compositions of the dust likely form a continuum, so that there are cases which are not clear-cut. The spectra are compared in Figure 3. Information about the circumstellar disks and comets used are in Table 3.

### 4.1. Grouping of Astronomical Spectra

Group 1 (Fig.3) represents simple spectra like GK Tau, dominated by an amorphous silicate feature at ~10 µm in the 8 – 13 µm range. Shoulders at ~11 µm are the only additional feature in this spectral region, indicating minor crystalline components (Sargent et al., 2009; Oliveira et al., 2011; Speck et al., 2011). In contrast, samples showing clear crystalline features, such as the 11.3 µm peak of olivine belong in group 3 (e.g. Morlok et al., 2006a).

In the 20 µm region, we see also weak crystalline enstatite and forsterite features emerging from ~19.5µm onwards, with a potential sulfide or forsterite band at ~23.5 µm (Keller et al., 2002; Sargent et al., 2009; Oliveira et al., 2011).

Group 1 spectra are similar to interstellar medium (ISM) spectra, which are regarded as a sign of very pristine, unprocessed dust (Kemper et al., 2004). However, group 1 spectra also occur in older T Tauri systems, even in transitional disks (Sargent et al., 2006; Watson et al., 2009).

Group 1 shows similarity with the spectra of type 2 matrix with medium shock stage in the bands at 9.9 µm/11.2 µm (also compare Morlok et al., 2010) or mesostasis-rich chondrules (10.1 µm) (Tab.2). This indicates that such amorphous materials could be also result of circumstellar processing, which should be expected in older disks after several million years. Watson et al. (2009) explain this with in-fall of pristine material from the outer disk parts e.g. by planet formation after the processed material was cleared in the inner system ('Born again Disk').

Group 2a (Fig.3) is characterized in the 10 µm region by a broad feature of amorphous pyroxene and olivine, as well as clear bands of crystalline forsterite, enstatite and silica (~9.2, ~10.1 and ~11.1 µm) in the Tauri systems V955 Tau, DK Tau and Herbig star HD104237 (Chihara et al., 2002; Sargent et al., 2009; Juhasz et al., 2010; Oliveira et al., 2011; Speck et al., 2011). In the 20 µm region, olivine and pyroxene bands at ~19, ~23 and ~33 µm dominate (Koike et al., 2000 and 2003). Smaller pyroxene features are seen at 12.5 and 13.5-14.5 µm (Chihara et al., 2002).

The crystalline features indicate significant processing, with the pyroxene-rich Kakangari (K3) matrix similar with bands at 9.3, 9.9, 10.2, and 11.1 µm. Pyroxene-rich rim material from the Ornans CO3.3 CAI is less similar. However, a spinel feature at ~14.4 µm is similar to a strong peak in V955 Tau. This would indicate the occurrence of CAI-type highly refractory phases, with a very low degree of alteration (Morlok et al., 2008). Spectra showing only a shoulder at ~9.2 µm are in group 3.

Group 2b (Fig.3) is a subgroup consisting of spectra with the 9.2 µm feature dominating the 8-13 µm range, pointing towards a composition with significant crystalline enstatite and possibly high silica contents e.g. for low mass star Oph 3 (Chihara et al., 2002; Bouwman et al., 2008; Speck et al., 2011). None of the single components in this study reproduces the strong characteristic 9.1-9.2 µm band, only the bulk of enstatite chondrite Indarch (EH4) is comparable to Oph 3, but also has additional features in the 10 µm region. In order to reproduce the 11.1 µm features the presence of chondrules or matrix is could be necessary.

Group 3a (Fig.3) is dominated in the 10 µm region by a strong forsterite feature at 11.2-11.3µm. Samples are transitional Herbig and T Tauri disks HD100546 and EK Cha. A strong band is visible at ~10 µm. Further peaks are at ~16.3, ~19.6, ~23 and 33 µm. This reflects the dominance of amorphous olivine and amorphous pyroxene in the inner disk, but with significant crystalline forsterite, enstatite and minor silica (Koike et al., 2003; Lisse et al., 2007; Sicilia-Aguilar et al., 2009; Juhasz et al., 2010). Olivine-rich chondrule material is similar to this type of spectra: The characteristic 11.3 µm band is reproduced, also the 10 µm and the emergent pyroxene feature in EK Cha at 9.3 µm. Recrystallized, olivine-rich material from shocked Murchison (CM2) matrix is a further suggested source (Morlok et al., 2010). There are also significant PAH features at 6.2, 7.7, 8.7 in HD100546 (Lisse et al., 2007).

Group 3b (Fig.3). This sub-group has a feature at ~10 µm with a comparable or higher intensity as the 11.3µm band. An example of the latter case is 17P/Holmes (Reach et al., 2010), with the strongest feature at 10µm, at 11.0 µm and 19.3 µm. An average of the spectra of Murchison (CM2) matrix from medium shock (21-34 GPa) provides a similar spectrum. A 1:1 mixture of olivine/mesostasis-rich chondrule material is also similar.

Comets 9P/Tempel1 (post-Deep Impact ejecta), Hale-Bopp (Crovisier et al., 1997; Lisse et al., 2007) and 73P/Schwassmann-Wachmann 3 (Sitko et al., 2011) show features of similar strength at ~10.0-10.2 µm, 10.9-11.2 µm. These are similar to a mixture of medium to high (21 – 36 GPa) shocked Murchison CM2 matrix (also compare findings in Morlok et al., 2010). Also, an even mixture of the average olivine-rich and mesostasis-rich chondrules provides a similar spectrum, except for the 10.9 µm band of 73P/Schwassmann-Wachmann 3 in both cases (Sitko et al., 2011). Comet 73P/Schwassmann-Wachmann 3 is also comparable to highly shocked Murchison CM2, but the band positions are more similar to the 21-36 GPa mixture.

In addition to the comparison of the spectra based on band positions and shape, the series of data for group 3 allows to quantify similarities by using normalized spectra of the region between 8 µm and 13 µm. This is especially of interest regarding the identification of chondrule materials. Based on the feature at ~11.3 µm (maximum intensity of olivine-rich chondrules) and the band at ~10.0 µm (maximum intensity of mesostasis–rich chondrules) we calculated the ratios $F_{1/2}=F_1/F_2$ for integrated intensities of the areas from 9.9 µm-10.3 µm ($F_1$) and 11.0 µm-11.4 µm ($F_2$) (Fig.5a).

Averages for olivine rich chondrules ($F_{1/2}$=0.85), mesostasis rich chondrules ($F_{1/2}$=1.53) and Type 3 matrix ($F_{1/2}$=0.71) indicate that the strength of the amorphous material that affects $F_1$ allows to distinguish between the groups. Bulk analyses for ordinary chondrites Parnallee ($F_{1/2}$=0.88), Ceniceros ($F_{1/2}$=0.89) and Baratta ($F_{1/2}$=0.88) are very similar to the ratio for the olivine-rich chondrules (Fig.5b). The low ratio for type 3 matrix could be a potential way to identify chondrules in dust spectra, although there is an overlap between the two sets: Olivine-rich chondrules range from $F_{1/2}$ =0.97 to 0.79, type 3 matrixes from $F_{1/2}$= 0.5 to 0.85.

Ratios for representative samples in group 3 cover a comparable range. HD144688 ($F_{1/2}$=0.88) and HD100546 ($F_{1/2}$=0.90) fall into the range of olivine-rich chondrules, and are also very similar to the bulk ordinary chondrites (Fig.5b). HD100546 shows some strong PAH features at 6.2, 7.7, 8.7 µm (Lisse et al., 2007), which probably do not influence the $F_1$ band significantly.

Comets 9P/Tempel1 (post-Deep Impact ejecta; $F_{1/2}$=0.99), Hale Bopp ($F_{1/2}$=1.03), 73P/Schwassmann/Wachmann ($F_{1/2}$=1.06; not shown in Figure 5) and 17P/Holmes ($F_{1/2}$=1.31; not shown in Figure 5) show the highest ratios, above the olivine-rich chondrule range and closer to the mesostasis rich chondrules ($F_{1/2}$=1.35-1.69). However, comets probably contain abundant other kinds of amorphous silicates like GEMS (Fig.5a, b).

The inner disk of HD142527 ($F_{1/2}$=0.53) falls into the type 3 matrix range ($F_{1/2}$=0.5-0.85), the ratio for EK Cha ($F_{1/2}$=0.82) into the overlapping range between olivine-rich and mesostasis-rich chondrules.

Group 4 (Fig.3) consists only of the complex spectrum of Herbig star HD179218. It shows strong enstatite features in the 10 µm range at 9.3, 9.9, 10.6, 11.2, 11.6, and ~12.5 µm (Chihara et al., 2002). The modeled mineralogy is dominated by amorphous silicates with significant enstatite and silica in the inner system (Juhasz et al., 2010; van Boekel et al., 2005). There are also significant PAH features at 6.2, 7.7, 8.7 and 11.3 µm (Lisse et al., 2007). A mixture of 50% K3 Kakangari matrix and 50% Acfer 207 matrix is very similar, with the characteristic 'twin-peak' at 10.6 and 11.2 micron, as well as the bands at 9.3 µm and 9.8 µm. It has shoulders at the position of the 11.6 µm feature. A mixture of 75% Kakangari matrix and 25% olivine-rich chondrule 11 also provides a very similar result.

We grouped previously published spectra using the same criteria as for group 1 to 4. Groups 1-3 occur all among Herbig and T Tauri systems. Among Herbig systems, group 1 and 2 are similarly abundant (36/38%), followed by group 3 (25%). The only sample of group 4 is HD179218 (N=61; Leinert et al., 2004; van Boekel et al., 2005; Juhasz et al.,2010; Voerhoeff et al., 2012). In T Tauri systems, group 1 (45%) dominates, followed by group 2 (29%/26%) (N=82; Watson et al., 2009). Also, groups 1-3 are typical for giants (Kastner et al., 2010) and low mass stars/brown dwarfs (Riaz, 2009; Sicilia-Aguilar et al., 2008). Such similarities have been observed in other studies (e.g. Sicilia-Aguilera et al., 2007; Oloffson et al., 2009).

### 4.2. Space/Time Resolved Spectra (Fig.4)

The space-resolved spectra of three Herbig systems in van Boekel et al. (2004) allow a more detailed insight into the compositional structure of a protoplanetary disk. They also provide possibly the 'cleanest' spectra for the different parts of a disk.

The inner 1-2 AU of transitional disk HD142527 show a group 2/3 spectrum, has a dominant forsterite band at 11.3 μm, but also significant pyroxene/enstatite contents as indicated by the peak at 9.3 μm and weak feature at 10.3 μm (Chihara et al., 2002; Koike et al., 2003; van Boekel et al., 2004). Porphyritic olivine and pyroxene-rich chondrule 11 shows very similar shaped bands and positions, as does bulk ordinary chondrite Baratta for the bands at 11.3 and 9.3 μm (Fig.4). However, when band ratios are taken into consideration, type 3 fine-grained matrixes show the biggest similarity (Fig.5a).

The outer disk (2-20 AU) has a group 2 spectrum with strong pyroxene features at 9.3, 10.0, 10.6 and 11.1 μm (Chihara et al., 2002). These are all similar to the Kakangari (K3) matrix. Van Boekel et al. (2004) also report slight crystalline pyroxene dominance in that part of the disk.

Inner disks of Herbig stars HD 163296 and HD 144432 are similar, again with a clear ~2:1 majority of crystalline forsterite in the modeled spectra (van Boekel et al., 2004). While they look like group 2 spectra they actually have slightly different band positions with broad features at ~9.5, ~10.3 and ~11.2 μm (Fig.4). Highly shocked Murchison CM2 matrix only shows a shoulder at the 9.5 μm position, but is similar to the 9.5 μm and 11.2 μm features of inner HD144432. The outer disks of HD 163296 and HD 144432, as well as EX Lupi pre-outburst are typical group 1 spectra with low crystallinity, and a feature at ~10µm (Abraham et al., 2009; van Boekel et al., 2004). The material is possibly still pristine but material processing (e.g. by shock) cannot completely ruled out.

Generally, in the inner parts of the disks there is a (weak) trend towards highly processed materials such as chondrules and (shocked) matrix, while the pristine outer parts are more difficult to relate to such solar system materials. Post-outburst EX Lupi has an additional broad band at ~10.5-11.1µm, while the 10 µm peak 'sharpens' slightly (Abraham et al., 2009), indicating material processing is forming crystalline features. Candidates here are mesostasis-rich chondrules and shocked Murchison (CM2) matrix for the 10 µm band and (barely) the emerging feature. Given the age of the system (3 Myr; Tab.3), both processes are possible if the timescales are comparable to our Solar System (e.g. Apai and Lauretta, 2010).

**5. Conclusions**

Mid-infrared spectra of chondrules, CAIs and fine-grained matrix material were compared to the petrology of the dust in young circumstellar disks and comets.

Four different major groups of spectra were identified in our analysis. Group 1 represents mainly amorphous silicate material, and shows similarity to processed components such as mesostasis-rich chondrules or highly shocked fine-grained matrix. Most of Group 2 is similar to heat-processed components - enstatite-rich matrix material from the Kakangari K3 chondrite. Group 3 has similarities to varying mixtures of olivine-rich and mesostasis-rich chondrule material, bulk ordinary chondrites and olivine-rich matrix material. Materials that show a strong amorphous 10 µm feature are comparable to shock-processed type 2 chondrite materials, but also mixtures of chondrule material. The feature rich spectrum HD179218 (group 4) requires a mixture of heat-processed materials such as chondrule and matrix.

In spatially resolved observations of circumstellar disks we find chondrule components and bulk ordinary chondrite material showing similarity to the 8-13 µm feature of the dust in the inner disks.

If the band intensities of the chondrite components are taken into account, group 3 materials can be explained by mixtures of olivine-rich, mesostasis-rich chondrules, fine-grained matrix and bulk ordinary chondrites.

However, for several spectra in group 3 also the impact processed type 2 carbonaceous chondrite material explains the features. Here the age of a system will be deciding to interpret the data.

Of CAI-type materials, only a potential spinel feature was found in several disks. Given the small modal abundance of CAI, this is not entirely surprising. This is also confirmed by earlier studies (Morlok et al., 2008). Bensity et al. (2010) and Suh et al. (2011a; 2011b) found corundum as minor highly refractory component in T Tauri and Herbig disks.

How can we link these findings to the evolution of Solar Systems? Generally, the basic similarity of dust in various young stellar environments to materials from our early Solar System indicates that they are possibly analog materials. The spectral groups 1-3 occur in all classes of systems and evolutionary stages of the disks, indicating that our Solar System is typical in the types of material produced around a newly-forming star. Mineral processing was observed both very early and late in protoplanetary disks, up to 1 to 8 Myr (Oliveira et al., 2011). The age range of these objects overlaps with that of chondrule formation in our Solar System (up to ~4 Myr; Apai and Lauretta, 2010). The presence of shocked type 1 and 2 materials is also possible with the early onset of aqueous alteration, at 4.4-5.7 Myr (Fujiya et al., 2013).

The timing of similar processes in other planetary systems could be different than in our early Solar System, but the probability of finding unaltered chondritic components is higher in young systems like e.g. HD142527 (1 Myr, Tab.3). With increasing age of a system (e.g. HD144432, 10 Myr, Tab.3), the probability of alteration increases. In this case, aqueous alteration and impacts could play a bigger role in the composition of the observed materials.

Based on models of material processing in an equilibrium process by evaporation/condensation, Gail (2004; 2010) observed a similarity between fine-grained matrix in primitive meteorites and the modeled dust composition. Similarities observed in this study between spectra of fine-grained matrix from type 3 chondrites and circumstellar dust would support these findings. However, the alternative possibility of highly processed meteoritic components from chondrule formation or impact-shocked material as major events cannot be discounted.

Extensive studies (e.g. Sicilia-Aguilar et al.,2007; Watson et al.,2009; Olofsson et al.,2010 and Oliveira

et al.,2011) have found few correlations between mineral compositions, crystallinity, disk chemistry, and age, mass or spectral types of the systems. Sargent et al. (2009) and Watson et al. (2009) found positive correlations between all crystalline components in circumstellar disks, pointing towards non-equilibrium processes. This indicates that large scale processes such as planet formation can erase most of such traces. This also makes it difficult to find any meaningful correlation between the meteorite components identified in this study with the range of disk characteristics of the discussed systems. Comets also show a variety in their spectral features, with many falling into group 3. This, however, could be an artifact due to the comparatively low number of available spectra.

Wooden (2008) find that comets are dominated by pre-accretionary materials. However, analyses of cometary grains from the Stardust mission show they are also similar to fragmented chondrules and CAIs (e.g. Ishii et al., 2008; Nakamura et al., 2008; Chi et al., 2009; Bridges et al., 2012). Comets in group 3 actually show similarities with chondrule and shock processed type 2 fine-grained matrix material.

This would allow another source of processed materials in addition to particles transported outwards from the inner solar system. Tomeoka et al. (2008) found evidence for impact shock pressure greater than 27 GPa in Stardust olivine. This is in the range of the shocked chondrite material (21-49 GPa). Changes in the mid-infrared properties of forsterite due to impact shock have been also observed by Lederer at al., 2012.

The source regions of comets, the Kuiper belt and the Oort cloud show low relative impact speeds (< 2 km $s^{-1}$). Still, short-period comets originating in the Kuiper belt may have been processed by continuous impacts (Durda and Stern, 2000).

Orbital instability of the giant planets orbits in the early Solar System (as proposed in the Nice model) could also have enabled collisions at higher relative speeds by scattered comets (e.g. Morbidelli et al., 2009). The question remains how substantial material processing of silicates still allows high volatile contents (e.g. CO, $CO_2$) in comets. However, the materials could have experienced high impact shock in the inner Solar System before being transported outwards and being incorporated into comets.

Further components to be taken into account are the primitive interplanetary dust particles (IDPs) and glass with embedded metal and sulfides (GEMS), which are expected to be present in the earliest stages of disk evolution. These materials already can exhibit variation in their mineralogy and spectral features (e.g. Brunetto et al., 2011; Davidson et al., 2012; Starkey et al., 2013). Especially the amorphous 'bump' at ~10 µm also could be created by 'pristine' amorphous materials. Interplanetary dust particles enriched with GEMS show amorphous features in the 10 µm area (Bradley et al., 1999). Unfortunately, the number of mid-infrared analyses of IDP and GEMS is limited due to technical difficulties based on the small particle size (usually <10 µm). Also, this small size makes the single IDP spectra probably less representative than a complete component, which sums up a larger cross section of materials, especially in bulk or matrix analyses.

IDP or GEMS material, and similar types of amorphous silicates in circumstellar disks, would make the identification of chondrule-like materials based on band ratios difficult. The feature of amorphous mesostasis in chondrules is at a similar position, so it would be difficult to differentiate between the two materials. For a complete comparison with dust in circumstellar environments and comets, a larger number of mid-infrared spectra from cometary sources with a high signal-to-noise ratio are needed, as well as laboratory spectra of IDP.

While it is difficult to confirm the occurrence of chondrule-like material with this complication, we could still extract information about chondrule formation from astronomical dust spectra. A low $F_{1/2}$ ratio and thus very low contents of amorphous silicates could point towards the absence or a low intensity of chondrule formation in such systems. A candidate would be the dust in the inner disk of the young (1 Myr) HD142527 system. The material falls clearly outside the range of the chondrule materials and is very similar to fine-grained matrix.

Finally, the suite of spectra from primitive materials is not complete; the high variety of materials also demands further investigations in order to complete the picture. This includes more studies of IDP and GEMS material, CAIs but also of chondrule mesostasis material.

Furthermore, more mid-infrared spectra of comets are needed. For example, there is scant mid-infrared data available for comet 81P/Wild 2 (Hanner and Hayward, 2003). In order to allow a better comparison between the detailed laboratory data of dust particles from the Stardust mission and spectroscopic observations of other comets, better mid-infrared data for 81P/Wild 2 would be highly desirable.

**Acknowledgements**

Many thanks to Roy van Boekel (Heidelberg) for the space resolved MIDI spectra, to Takahito Osawa (Ibaraki) for the additional matrix data. Also, many thanks to Caroline Smith (London) for providing the Allende samples.

This work is partially based on observations made with the Spitzer Space Telescope, which is operated by the Jet Propulsion Laboratory, California Institute of Technology, under NASA Contract 1407. Part of this work is based on observations with ISO, an ESA project with instruments funded by ESA Member States and with the participation of ISAS and NASA. Also, part of the data is based on observations obtained at the European Southern Observatory (ESO).

This research has made use of the SIMBAD database, operated at CDS, Strasbourg, France.

**Figure Captions:**

Figure 1: BSE images of representative areas in meteorites sampled for the infrared analyses in this study. A-C: Chondrules (a: PO chondrule 1; b: BO chondrule 7; c: RP chondrule 8); D-I: Fine-grained matrix. J-L: CAIs (based on Morlok et al., 2008).

Figure 2: Laboratory infrared spectra of chondrules, CAI and matrix material. In absorbance units. For literature sources see text

Figure 3: Comparison of laboratory data with astronomical spectra of circumstellar dust and comets (in absorbance units; micron). Black spectra: Astronomical data in Flux ($F_\nu$), Grey: laboratory data. The spectra are divided into 4 groups depending on common spectral characteristics. For source of astronomical spectra see text.

Figure 4: Comparison of laboratory data (grey, in absorbance units) with astronomical spectra of circumstellar dust in space- resolved spectra of three Herbig systems (van Boekel et al., 2004) and time – resolved EX Lupi (Abraham et al., 2009) (black, in Flux ($F_\nu$)).

Figure 5: Comparison of band intensities at 9.9-10.3 μm and 11.0-11.4 μm normalized spectra for (a) averages of olivine and mesostasis-rich chondrules and olivine-rich fine-grained matrix and (b) bulk analyses of ordinary chondrites Parnallee LL3.6, Baratta L3.8 and Ceniceros H3.7 (in absorbance).

The laboratory spectra are compared to astronomical spectra (in $(F_\nu)/(B_\nu(T))$ of dust from EK Cha (Silica-Aguilera et al., 2009), HD144688 (van Boekel et al., 2005), the inner disk of HD142527 (van Boekel et al., 2004), HD100546, comets Hale Bopp and post-Deep Impact ejecta of 9P/Tempel1 (Lisse et al., 2007).The grey fields mark the areas integrated to calculate the $F_1/F_2$ ratios.

| | Chondrule 1 (PO) | Chondrule 2 (PO) | Chondrule 4 (PO) | Chondrule 9 (PO) | Chondrule 10 (PO) | Chondrule 7 (BO) | Chondrule 3 (POP) | Chondrule 5 (POP) | Chondrule 11 (POP) | Chondrule 8 (RP) | Chondrule 6 (PP) |
|---|---|---|---|---|---|---|---|---|---|---|---|
| **Olivine** | | | | | | | | | | | |
| $Na_2O$ | n.d. | n.d. | n.d. | n.d. | 0.04±0.14 | n.d. | n.d. | n.d. | n.d. | n.d. | |
| MgO | 55.52±0.47 | 53.45±0.52 | 54.18±1.55 | 44.07±9.94 | 51.56±4.08 | 56.47±0.53 | 52.86±1.23 | 50.68±1.69 | 53.86±1.43 | 29.49±0.65 | |
| $Al_2O_3$ | n.d. | n.d. | n.d. | n.d. | 0.74±2.25 | n.d. | n.d. | n.d. | n.d. | 0.16±0.27 | |
| $SiO_2$ | 42.18±0.24 | 41.12±0.07 | 41.66±0.52 | 39.43±2.21 | 41.56±0.58 | 42.35±0.34 | 41.59±0.43 | 41.14±0.66 | 41.43±0.60 | 36.07±0.44 | |
| CaO | 0.45±0.14 | 0.10±0.14 | 0.38±0.22 | 0.10±0.14 | 0.50±0.54 | 0.09±0.16 | 0.13±0.13 | n.d. | 0.36±0.27 | n.d. | |
| $Cr_2O_3$ | n.d. | n.d. | n.d. | 0.17±0.24 | 0.14±0.27 | 0.11±0.19 | n.d. | n.d. | 0.07±0.15 | n.d. | |
| FeO | 1.24±0.74 | 4.02±0.46 | 3.55±2.22 | 15.63±12.51 | 5.69±2.29 | 0.82±0.15 | 4.78±2.04 | 7.95±2.05 | 3.65±1.45 | 33.72±0.86 | |
| Total | 99.40±0.73 | 98.68±0.21 | 99.76±0.65 | 99.39±0.75 | 100.24±0.94 | 99.84±1.03 | 99.36±0.61 | 99.76±0.46 | 99.38±0.51 | 99.44±0.08 | |
| N | 6 | 3 | 5 | 6 | 15 | 4 | 4 | 6 | 5 | 4 | |
| **Pyroxene** | | | | | | | | | | | |
| MgO | | | | | | | 38.25±0.03 | 38.52±0.31 | 37.75±0.16 | 32.24 | 33.92±0.31 |
| $Al_2O_3$ | | | | | | | 0.64±0.39 | 0.61±0.06 | 0.59±0.08 | 4.38 | 11.08±0.56 |
| $SiO_2$ | | | | | | | 59.01±0.52 | 59.12±0.34 | 59.02±0.42 | 51.37 | 51.54±0.45 |
| CaO | | | | | | | 0.68±0.11 | 0.59±0.10 | 0.64±0.06 | 4.00 | 0.55±0.06 |
| $Cr_2O_3$ | | | | | | | 0.61±0.14 | 0.57±0.04 | 0.62±0.05 | 0.87 | 0.60±0.12 |
| FeO | | | | | | | 0.61±0.41 | 0.72±0.23 | 0.80±0.20 | 5.18 | 0.26±0.27 |
| $TiO_2$ | | | | | | | n.d. | n.d. | n.d. | 0.43 | 0.96±0.13 |
| Total | | | | | | | 99.80 | 100.01 | 99.41 | 98.47 | 99.17 |
| N | | | | | | | 4 | 5 | 2 | 1 | 6 |
| **Mesostasis** | | | | | | | | | | | |
| $Na_2O$ | 1.07±0.12 | | | | 1.92 | 1.74±0.31 | | | | | 1.70 |
| MgO | 9.75±1.22 | | | | 5.13 | 5.54±0.88 | | | | | 0.48 |
| $Al_2O_3$ | 23.53±0.22 | | | | 25.39 | 22.34±0.68 | | | | | 32.08 |
| $SiO_2$ | 47.08±1.28 | | | | 48.96 | 47.47±0.62 | | | | | 47.06 |
| $SO_3$ | 0.7±0.7 | | | | n.d. | n.d. | | | | | n.d. |
| CaO | 16.62±1.15 | | | | 14.92 | 19.31±0.68 | | | | | 16.97 |
| FeO | 1.47±0.94 | | | | 1.78 | 0.79±0.03 | | | | | n.d. |
| $TiO_2$ | 0.81±0.01 | | | | 0.59 | 1.05±0.01 | | | | | n.d. |
| Total | 101.02 | | | | 98.69 | 98.22 | | | | | 98.29 |
| N | 2 | | | | 1 | 2 | | | | | 1 |

Table 1a. Analyses of minerals/components in the chondrules used for infrared analyses, with standard deviation (in wt%). N=Number of analyzed mineral grains. N.d.=not detected/below detection limit of 0.02wt%. For explanation of chondrule types see text.

| | Alais C1 | Al Rais CR2 | Cold Bokkeveld CM2 | Ningqiang CK3 | Ornans CO3.3 | Vigarano 1/2 CV3.3 | Kakangari K3 |
|---|---|---|---|---|---|---|---|
| $Na_2O$ | 0.25±0.22 | 0.58±0.19 | 0.45±0.19 | 1.29±1.47 | 0.26±0.40 | 0.62±0.52 | 1.14±0.05 |
| MgO | 16.53±0.36 | 15.76±1.68 | 19.55±0.72 | 16.18±2.87 | 15.46±1.86 | 15.85±1.25 | 27.23±0.54 |
| $Al_2O_3$ | 1.88±0.08 | 2.03±0.69 | 1.65±0.68 | 5.61±3.68 | 2.25±0.53 | 3.20±0.81 | 2.66±0.56 |
| $SiO_2$ | 27.66±0.63 | 28.06±2.51 | 30.12±0.88 | 30.89±2.28 | 24.45±2.73 | 28.46±1.04 | 40.73±5.79 |
| $P_2O_5$ | n.d. | 0.23±0.34 | 0.46±0.30 | 0.02±0.08 | 0.08±0.29 | 0.02±0.10 | n.d |
| $SO_3$ | 3.49±0.46 | 8.17±2.37 | 6.83±0.88 | 0.47±0.39 | 1.31±2.22 | 1.11±0.43 | 6.21±3.37 |
| $K_2O$ | n.d | n.d | n.d | 0.12±0.17 | n.d | n.d. | 0.12±0.11 |
| CaO | 0.36±0.13 | 1.35±0.88 | 0.97±0.13 | 2.05±1.72 | 1.05±0.77 | 2.02±0.98 | 1.35±0.16 |
| $Cr_2O_3$ | 0.47±0.09 | 0.26±0.24 | 0.38±0.16 | 0.34±0.17 | 0.40±0.22 | 0.38±0.15 | 0.52±0.13 |
| MnO | n.d | 0.05±0.12 | 0.04±0.10 | 0.11±0.18 | 0.08±0.16 | 0.08±0.17 | 0.25±0.21 |
| FeO | 20.90±0.72 | 22.45±2.71 | 22.70±1.18 | 34.26±5.28 | 30.06±3.29 | 38.70±2.45 | 13.21±4.42 |
| NiO | 1.12±0.01 | 1.88±0.43 | 1.39±0.10 | 0.08±0.22 | 0.86±0.31 | 1.67±0.42 | 0.76±0.16 |
| $TiO_2$ | n.d | 0.01±0.05 | n.d. | 0.05±0.13 | n.d | n.d | n.d |
| | 72.63±1.26 | 80.83±4.49 | 84.52±1.57 | 91.47±3.22 | 76.26±5.26 | 92.11±1.21 | 94.17±1.05 |

Table 1b. Representative EDX analyses of fine-grained matrix areas in chondrites (in wt%). Rim areas of chondrules were avoided.

| | Ornans (CO3.3) Spinel (INT) | Gehlenite (INT) | Fassaite (RIM) | Diopside (RIM) | Vigarano 1 (CV3.3) Spinel (INT) | Gehlenite (INT) | Perowskite (INT) | Vigarano 2 (CV3.3) Diopside (RIM) | Forsterite (RIM) | Spinel (INT) | Gehlenite (INT) | Anorthitic (RIM) |
|---|---|---|---|---|---|---|---|---|---|---|---|---|
| **$Na_2O$** | n.d. | n.d. | n.d. | n.d. | n.d. | n.d. | n.d. | n.d. | n.d. | n.d. | n.d. | 10.50 |
| **MgO** | 28.10 | 2.70 | 14.30 | 19.30 | 28.60 | n.d. | n.d. | 19.40 | 55.90 | 29.50 | 4.80 | 3.30 |
| **$Al_2O_3$** | 68.90 | 29.60 | 14.30 | 0.60 | 68.80 | 35.20 | 0.47 | n.d. | n.d. | 70.50 | 24.70 | 31.20 |
| **$SiO_2$** | n.d. | 26.90 | 46.50 | 56.20 | n.d. | 22.80 | n.d. | 55.80 | 43.70 | n.d. | 29.50 | 43.40 |
| **$V_2O_5$** | 0.51 | n.d. | n.d. | n.d. | 0.98 | n.d. | n.d. | n.d. | n.d. | n.d. | n.d. | n.d. |
| **K2O** | n.d. | n.d. | n.d. | n.d. | n.d. | n.d. | n.d. | n.d. | n.d. | n.d. | n.d. | 0.37 |
| **CaO** | n.d. | 41.10 | 24.70 | 24.70 | 0.33 | 41.40 | 40.80 | 25.20 | 0.60 | n.d. | 41.40 | 9.00 |
| **FeO** | 1.05 | n.d. | 0.46 | n.d. | 0.89 | n.d. | n.d. | 0.78 | 1.30 | 0.57 | n.d. | 1.50 |
| **$TiO_2$** | n.d. | n.d. | 0.46 | n.d. | n.d. | n.d. | 58.20 | n.d. | n.d. | n.d. | n.d. | n.d. |
| **Total** | 98.50 | 100.30 | 100.60 | 100.80 | 99.60 | 99.30 | 99.50 | 101.20 | 101.50 | 100.50 | 100.40 | 99.30 |

Table 1c. Representative EDX analyses of characteristic phases in the Calcium-Aluminium-rich Inclusions (CAI) used in this study (based on Morlok et al., 2008). INT=Interior part of CAI, RIM=Outer part. Anorthitic: Phase with composition similar to anorthite. In wt%.

| Band | Barycenter | FWHM | Int. | Band | Barycenter | FWHM | Int. | Band | Barycenter | FWHM | Int. |
|---|---|---|---|---|---|---|---|---|---|---|---|
| **Chon1** | **(PO)** | | | **Chon10** | **(PO)** | | | **Chon5** | **(POP)** | | |
| 9.37 | 9.32±0.09 | 2.01±0.16 | 1.18 | 9.35 | 9.24±0.01 | 0.79±0.02 | 0.43 | 9.42 | 9.34±0.01 | 0.75±0.03 | 0.37 |
| 10.14 | 10.14±0.04 | 1.38±0.24 | 0.67 | 10.17 | 10.09±0.01 | 0.78±0.03 | 0.90 | 10.11 | 10.02±0.02 | 0.57±0.03 | 0.55 |
| 11.30 | 11.30±0.01 | 1.70±0.06 | 2.33 | 10.60 | 10.79±0.02 | 0.76±0.09 | 0.45 | 10.44 | 10.55±0.02 | 0.73±0.12 | 0.32 |
| 11.95 | | | | 11.30 | 11.32±0.01 | 0.68±0.03 | 0.78 | 11.22 | 11.22±0.01 | 0.87±0.03 | 1.12 |
| 14.97 | | | | 11.90 | 11.98±0.01 | 0.37±0.03 | 0.11 | 11.88 | 11.91±0.01 | 0.18±0.05 | 0.03 |
| 16.56 | 16.41±0.02 | 1.37±0.12 | 0.77 | 13.60 | | | | 13.80 | | | |
| 19.76 | | | | 14.97 | | | | 16.37 | 16.29±0.02 | 0.62±0.05 | 0.20 |
| 21.19 | 21.30±0.08 | 1.11±0.52 | 0.23 | 16.61 | 16.55±0.02 | 0.65±0.05 | 0.19 | 19.65 | 19.35±0.02 | 3.27±0.06 | 3.58 |
| 24.04 | 23.99±2.41 | 3.50±7.95 | 3.35 | 19.57 | 19.33±0.02 | 2.33±0.05 | 2.37 | 21.65 | 21.66±0.03 | 0.60±0.12 | 0.14 |
| 27.78 | 27.84±1.87 | 5.30±2.03 | 5.23 | 21.19 | 21.22±0.03 | 1.41±0.15 | 0.45 | 23.58 | 23.62±0.03 | 2.60±0.14 | 2.86 |
| 34.01 | 34.03±0.06 | 2.26±0.25 | 1.21 | 24.10 | 24.14±0.02 | 3.57±0.11 | 3.78 | 26.18 | | | |
| | | | | 27.86 | 28.37±0.05 | 4.18±0.20 | 2.61 | 27.62 | 27.69±0.05 | 3.46±0.18 | 1.96 |
| **Chon2** | **(PO)** | | | 34.01 | 34.65±0.06 | 5.18±0.23 | 2.48 | 31.35 | | | |
| 10.14 | 10.00±0.01 | 1.26±0.02 | 1.33 | | | | | 33.78 | 33.97±0.05 | 3.73±0.17 | 1.95 |
| 10.44 | 10.91±0.09 | 1.31±0.13 | 0.58 | **Chon7** | **(BO)** | | | | | | |
| 11.25 | 11.28±0.01 | 0.60±0.04 | 0.59 | 9.31sh | 9.22±0.01 | 0.60±0.04 | 0.15 | **Chon11** | **(POP)** | | |
| 11.92 | 11.98±0.01 | 0.23±0.04 | 0.05 | 10.17 | 10.06±0.00 | 0.74±0.02 | 0.78 | 9.35 | 9.16±0.01 | 0.38±0.03 | 0.12 |
| 16.37 | 16.35±0.01 | 0.62±0.04 | 0.25 | 10.47 | 10.65±0.02 | 0.37±0.07 | 0.06 | 10.17 | 10.14±0.01 | 1.33±0.03 | 1.41 |
| 19.72 | 19.28±0.01 | 2.38±0.07 | 2.35 | 11.25 | 11.23±0.00 | 0.85±0.01 | 1.20 | 10.86 | | | |
| 21.28 | 21.46±0.03 | 1.09±0.13 | 0.34 | 11.92 | | | | 11.25 | 11.26±0.01 | 0.95±0.02 | 1.19 |
| 23.87 | 23.65±0.05 | 2.48±0.18 | 2.75 | 16.42 | 16.39±0.01 | 0.51±0.03 | 0.19 | 13.59 | | | |
| 27.55 | 27.83±0.05 | 2.84±0.33 | 1.43 | 19.72 | 19.35±0.01 | 2.41±0.05 | 2.16 | 13.83 | | | |
| | | | | 21.23 | | | | 14.68 | | | |
| **Chon4** | **(PO)** | | | 23.87 | 23.71±0.05 | 2.82±0.11 | 2.96 | 15.50 | | | |
| 10.07 | 10.05±0.01 | 0.99±0.01 | 0.97 | 27.62 | 27.94±0.03 | 2.54±0.15 | 1.26 | 16.50 | | | |
| 10.47 | | | | 33.90 | 34.03±0.04 | 2.92±0.19 | 1.48 | 19.65 | 19.40±0.02 | 2.63±0.05 | 3.06 |
| 11.25 | 11.21±0.00 | 0.80±0.01 | 1.09 | | | | | 20.75 | | | |
| 11.92 | | | | **Chon3** | **(POP)** | | | 21.74 | | | |
| 16.42 | | | | 9.35 | 9.26±0.01 | 0.56±0.02 | 0.37 | 23.81 | 23.84±0.03 | 3.59±0.13 | 3.59 |
| 19.61 | 19.35±0.02 | 2.77±0.05 | 2.59 | 10.17 | 10.03±0.01 | 0.67±0.03 | 0.68 | 26.39 | 26.40±0.22 | 1.19±0.99 | 0.09 |
| 21.19 | | | | 10.44 | 10.58±0.02 | 0.85±0.12 | 0.49 | 27.86 | 28.31±0.07 | 2.69±0.21 | 1.53 |
| 23.87 | 23.86±0.02 | 3.06±0.08 | 3.15 | 11.25 | 11.26±0.01 | 0.90±0.02 | 1.11 | 33.90 | 34.23±0.08 | 3.59±0.25 | 1.46 |
| 27.62 | 27.92±0.05 | 2.92±0.19 | 1.24 | 11.88sh | | | | | | | |
| 33.78 | 34.06±0.06 | 3.42±0.19 | 1.41 | 13.62 | | | | **Chon8** | **(RP)** | | |
| | | | | 13.83 | | | | 9.51 | | | |
| **Chon9** | **(PO)** | | | 14.68 | | | | 10.04 | 10.04±0.01 | 1.34±0.02 | 2.00 |
| 10.01 | 10.01±0.01 | 1.17±0.02 | 1.68 | 16.42 | 16.34±0.02 | 0.93±0.07 | 0.28 | 11.25 | 11.27±0.01 | 0.76±0.03 | 0.61 |
| 11.25 | 11.25±0.01 | 0.62±0.04 | 0.36 | 19.57 | 19.27±0.02 | 2.62±0.06 | 2.87 | 11.88 | | | |
| 11.92 | | | | 21.01 | 21.31±0.07 | 1.82±0.35 | 0.44 | 13.87 | | | |
| 13.18 | | | | 23.87 | 23.81±0.03 | 3.32±0.13 | 3.40 | 16.50 | 16.61±0.05 | 1.32±0.21 | 0.32 |
| 14.97 | | | | 27.62 | 27.91±0.06 | 3.74±0.22 | 1.82 | 19.76 | 19.62±0.07 | 3.05±0.24 | 2.92 |
| 16.48 | 16.30±0.01 | 0.52±0.05 | 0.32 | 33.78 | 34.03±0.06 | 3.14±0.19 | 1.17 | 21.46 | 21.44±0.04 | 0.29±0.17 | 0.06 |
| 19.80 | 19.26±0.10 | 3.00±0.27 | 1.71 | | | | | 24.04 | 24.01±0.10 | 1.32±0.51 | 0.58 |
| 21.55 | 21.61±0.10 | 2.33±0.49 | 1.41 | | | | | 27.55 | 27.92±0.14 | 4.58±0.60 | 2.62 |
| 23.81 | 23.86±0.17 | 4.11±0.74 | 3.29 | | | | | 33.90 | 34.11±0.19 | 5.89±0.65 | 2.35 |
| 27.70 | 28.04±0.23 | 4.53±1.00 | 1.72 | | | | | | | | |
| 33.78 | 34.26±0.19 | 5.29±0.68 | 1.99 | | | | | | | | |

| Band | Barycenter | FWHM | Int. | Band | Barycenter | FWHM | Int. | Band | Barycenter | FWHM | Int. |
|---|---|---|---|---|---|---|---|---|---|---|---|
| **Chon6** | **(PP)** | | | **Parnallee** | **(LL3.6)** | | | **Indarch** | **(EH4)** | | |
| 9.35sh | 9.36±0.01 | 1.27±0.03 | 0.74 | 9.51 | 9.30±0.01 | 1.15±0.03 | 0.79 | 9.16 | | | |
| 10.20 | 10.21±0.00 | 0.88±0.02 | 1.02 | 10.28 | 10.19±0.01 | 1.10±9.05 | 0.90 | 9.35 | 9.24±0.01 | 1.58±0.02 | 2.31 |
| 11.12 | 11.18±0.01 | 1.11±0.03 | 0.72 | 10.56 | | | | 9.84 | 9.91±0.03 | 0.28±0.13 | 0.02 |
| 11.90sh | | | | 11.30 | 11.30±0.01 | 1.15±0.02 | 1.48 | 10.28 | | | |
| 13.59 | 13.58±0.01 | 0.12±0.02 | 0.03 | 11.92 | | | | 10.66 | 10.67±0.01 | 0.98±0.05 | 0.73 |
| 14.03 | 14.05±0.02 | 0.42±0.08 | 0.04 | 13.57 | | | | 11.09 | | | |
| 14.97 | | | | 14.73 | | | | 11.68 | 11.69±0.01 | 0.71±0.05 | 0.29 |
| 19.53 | 19.55±0.15 | 1.96±0.35 | 1.00 | 16.89 | | | | 12.67 | | | |
| 21.41 | 21.41±0.02 | 1.48±0.09 | 0.95 | 19.65 | 19.63±0.01 | 3.09±0.04 | 3.43 | 13.59 | | | |
| 22.88 | 22.88±0.01 | 0.75±0.07 | 0.34 | 24.51 | 24.34±0.22 | 3.16±0.09 | 2.62 | 13.83 | | | |
| 24.33 | 24.37±0.05 | 2.99±0.24 | 1.48 | 28.41 | 28.67±0.04 | 4.21±0.18 | 2.45 | 14.43 | | | |
| 28.33 | 28.31±0.07 | 4.23±0.30 | 1.51 | 34.36 | 34.65±0.09 | 2.99±0.31 | 0.66 | 14.68 | | | |
| | | | | | | | | 15.41 | | | |
| **Olivine rich** | | | | **Ceniceros** | **(H3.7)** | | | 17.58 | 17.57±0.05 | 1.51±0.09 | 0.74 |
| 9.41 | 9.23±0.02 | 0.49±0.06 | 0.19 | 9.54 | 9.42±0.11 | 2.06±0.09 | 1.62 | 18.45 | 18.46±0.04 | 0.65±0.17 | 0.18 |
| 10.15 | 10.10±0.01 | 0.97±0.05 | 1.01 | 10.24 | 10.32±0.91 | 1.44±1.64 | 0.74 | 19.49 | 19.51±0.02 | 1.42±0.13 | 1.11 |
| 10.46 | | | | 10.56 | 10.70±1.71 | 1.42±3.03 | 0.38 | 20.75 | 20.76±0.03 | 0.77±0.14 | 0.31 |
| 11.28 | 11.21±0.01 | 0.93±0.02 | 1.22 | 11.30 | 11.31±0.01 | 0.65±0.06 | 0.53 | 21.74 | 21.72±0.04 | 0.90±0.22 | 0.29 |
| 11.89 | | | | 11.95 | 11.97±0.04 | 1.09±0.13 | 0.58 | 23.09 | 23.12±0.06 | 2.22±0.23 | 1.29 |
| 16.45 | 16.41±0.02 | 0.60±0.06 | 0.17 | 13.59 | | | | 26.46 | 26.48±0.10 | 2.65±0.43 | 0.98 |
| 19.68 | 19.37±0.01 | 2.49±0.05 | 2.48 | 14.75 | | | | 28.17 | | | |
| 21.11 | | | | 16.92 | 16.76±0.02 | 0.83±0.08 | 0.23 | 28.99 | 28.96±0.13 | 2.72±0.33 | 1.02 |
| 23.84 | 23.89±0.02 | 3.76±0.08 | 4.02 | 20.08 | | | | 33.11 | | | |
| 27.68 | 28.09±0.03 | 3.03±0.13 | 1.47 | 24.39 | 24.32±0.03 | 3.08±0.14 | 2.38 | 33.90 | | | |
| 33.87 | 34.16±0.03 | 3.58±0.11 | 1.61 | 28.70 | 28.70±0.06 | 4.13±0.25 | 1.86 | 35.71 | 35.78±0.19 | 2.25±0.57 | 2.84 |
| **Mesostasis rich** | | | | **Baratta** | **(L3.8)** | | | | | | |
| 9.40sh | | | | 9.45 | 9.30±0.02 | 1.36±0.03 | 1.01 | | | | |
| 10.12 | 10.05±0.01 | 1.36±0.02 | 2.00 | 10.28 | 10.25±0.02 | 1.21±0.11 | 0.75 | | | | |
| 11.25 | 11.27±0.01 | 0.66±0.04 | 0.42 | 10.56 | | | | | | | |
| 11.89sh | | | | 10.86 | | | | | | | |
| 13.57 | | | | 11.30 | 11.34±0.01 | 1.39±0.03 | 1.75 | | | | |
| 16.50 | 16.37±0.02 | 0.49±0.08 | 0.11 | 13.59 | | | | | | | |
| 19.73 | 19.34±0.04 | 3.20±0.12 | 2.43 | 14.75 | | | | | | | |
| 21.46 | 21.39±0.03 | 1.24±0.14 | 0.62 | 16.90 | 16.84±0.02 | 0.62±0.07 | 0.15 | | | | |
| 22.81 | 22.69±0.05 | 1.61±0.30 | 0.72 | 24.39 | 24.28±0.03 | 3.32±0.12 | 2.70 | | | | |
| 27.76 | 28.60±0.07 | 3.54±0.30 | 1.24 | 28.50 | 28.57±0.06 | 3.50±0.19 | 1.62 | | | | |
| 33.92 | 34.30±0.07 | 5.29±0.24 | 1.69 | | | | | | | | |

| Band | Barycenter | FWHM | Int. | Band | Barycenter | FWHM | Int. | Band | Barycenter | FWHM | Int. |
|---|---|---|---|---|---|---|---|---|---|---|---|
| **Alais** | **(CI1)** | | | **Vigarano2** | **(CV3.3)** | | | **Kakangari** | **(K3)** | | |
| 6.11 | 6.23±0.05 | 0.29±0.17 | 0.05 | 9.29 | 9.23±0.02 | 0.35±0.06 | 0.13 | 9.29 | 9.24±0.01 | 0.52±0.03 | 0.62 |
| 6.89 | | | | 10.08 | 10.07±0.01 | 0.76±0.04 | 0.83 | 9.88 | 9.85±0.02 | 0.42±0.08 | 0.37 |
| 9.88 | 9.95±0.02 | 1.19±0.05 | 1.86 | 11.26 | 11.23±0.01 | 0.77±0.03 | 1.07 | 10.20 | 10.22±0.04 | 0.35±0.21 | 0.17 |
| 15.24 | 15.47±0.23 | 1.42±0.69 | 0.26 | 11.96 | | | | 10.59 | 10.61±0.03 | 0.55±0.17 | 0.56 |
| | | | | 13.59 | 13.60±0.09 | 0.41±0.32 | 0.05 | 11.11 | 11.12±0.02 | 0.51±0.13 | 0.55 |
| **Cold Bokkeveld** | **(CM2**) | | | 14.30 | 14.29±0.10 | 0.74±0.46 | 0.15 | 11.52 | 11.58±0.03 | 0.43±0.10 | 0.25 |
| 9.96 | 10.02±0.02 | 1.22±0.06 | 1.96 | 14.90 | 14.89±0.07 | 0.30±0.28 | 0.04 | 13.74 | | | |
| 15.82 | 15.79±1.08 | 2.73±3.12 | 0.26 | | | | | 15.53 | | | |
| **Al Rais** | **(CR2)** | | | **Olivine Rich Matrix** | | | | 16.03 | | | |
| 6.16 | 6.15±0.02 | 0.38±0.07 | 0.05 | 9.30 | 9.23±0.01 | 0.19±0.03 | 0.02 | | | | |
| 6.89 | 6.86±0.04 | 0.25±0.11 | 0.02 | 10.15 | 10.12±0.00 | 0.73±0.01 | 0.62 | | | | |
| 7.94 | | | | 11.31 | 11.26±0.00 | 1.06±0.01 | 1.54 | | | | |
| 9.80 | 9.93±0.01 | 1.09±0.02 | 1.68 | 11.97 | | | | | | | |
| 15.33 | 15.49±0.10 | 1.08±0.28 | 0.16 | | | | | | | | |
| | | | | **Gjuba** | **(CB3)** | | | | | | |
| | | | | 9.34 | 9.29±0.01 | 0.46±0.02 | 0.28 | | | | |
| | | | | 10.45 | | | | | | | |
| **Ninqiang** | **(CK3)** | | | | | | | | | | |
| 9.33 | 9.92±0.02 | 0.22±0.06 | 0.05 | 10.62 | 10.55±0.01 | 0.74±0.05 | 0.56 | | | | |
| 10.25 | 10.16±0.01 | 0.67±0.04 | 0.65 | 11.22 | 11.31±0.01 | 0.76±0.03 | 0.57 | | | | |
| 11.36 | 11.37±0.01 | 1.11±0.06 | 1.65 | 18.08 | 17.85±0.01 | 4.03±0.05 | 5.83 | | | | |
| 12.02 | 12.16±0.08 | 0.56±0.32 | 0.08 | 20.79 | | | | | | | |
| 14.05 | 14.08±0.07 | 1.00±0.20 | 0.19 | 21.83 | | | | | | | |
| | | | | 23.59 | 23.49±0.04 | 5.44±0.14 | 4.51 | | | | |
| **DHO015** | **(CK3)** | | | | | | | | | | |
| 9.3sh | 9.28±0.00 | 0.06±0.01 | 0.01 | **Acfer 207** | **(CH3)** | | | | | | |
| 10.31 | 10.17±0.00 | 0.90±0.10 | 0.86 | 9.35 | 9.22±0.01 | 0.61±0.05 | 0.44 | | | | |
| 11.22 | 11.16±0.00 | 0.90±0.06 | 1.33 | 10.17 | 10.00±0.05 | 0.88±0.26 | 0.41 | | | | |
| 11.95 | | | | 10.45 | | | | | | | |
| | | | | 10.70 | 10.71±0.10 | 1.84±0.08 | 2.28 | | | | |
| **Ornans** | **(CO3.3)** | | | 11.22 | 11.39±0.02 | 0.62±0.12 | 0.34 | | | | |
| 10.12 | 10.05±0.01 | 0.25±0.03 | 0.13 | 13.83 | | | | | | | |
| 10.46 | 10.35±0.02 | 0.33±0.07 | 0.11 | 14.70 | | | | | | | |
| 11.31 | 11.29±0.01 | 0.97±0.03 | 1.45 | 16.42 | 16.64±0.14 | 1.37±0.57 | 0.21 | | | | |
| 12.02 | 12.03±0.01 | 0.16±0.04 | 0.05 | 19.72 | 19.49±0.24 | 3.18±0.71 | 3.03 | | | | |
| | | | | 20.83 | 21.01±2.36 | 4.54±2.16 | 1.00 | | | | |
| **Vigarano1** | **(CV3.3)** | | | 23.70 | 23.71±0.09 | 2.64±0.32 | 2.11 | | | | |
| 10.12 | 10.12±0.01 | 0.62±0.03 | 0.52 | | | | | | | | |
| 10.42 | | | | | | | | | | | |
| 11.31 | 11.25±0.01 | 0.94±0.02 | 1.48 | | | | | | | | |
| 12.02 | | | | | | | | | | | |

| Band | Barycenter | FWHM | Int. | Band | Barycenter | FWHM | Int. | Band | Barycenter | FWHM | Int. |
|---|---|---|---|---|---|---|---|---|---|---|---|
| **Murchision** | **(CM2)** | | | **Ornans INT** | **(CO3.3)** | | | **Vigarano CAI2 RIM** | **(CV3.3)** | | |
| **Unshocked** | | | | 9.29 | 9.30±0.02 | 0.29±0.07 | 0.24 | 9.47 | | | |
| 6.12 | | | | 10.29 | 10.10±0.10 | 1.12±0.46 | 0.89 | 9.69 | 9.78±0.02 | 0.86±0.05 | 0.68 |
| 7.05 | | | | 10.92 | | | | 9.84 | | | |
| 10.09 | 10.09±0.02 | 1.39±0.04 | 2.26 | 11.21 | 11.10±0.14 | 1.56±0.72 | 1.45 | 10.25 | 10.25±0.01 | 0.08±0.01 | 0.04 |
| 15.39 | 15.42±0.12 | 1.87±0.51 | 0.46 | 12.50 | 12.51±0.18 | 1.69±0.78 | 1.04 | 10.87 | 10.81±0.03 | 1.00±0.12 | 0.55 |
| 22.16 | | | | 14.54 | 14.44±0.04 | 1.31±0.14 | 1.86 | 11.11 | | | |
| | | | | | | | | 11.68 | | | |
| **Medium Shocked** | | | | **Ornans RIM** | **(CO3.3)** | | | 12.32 | 12.34±0.02 | 1.24±0.07 | 1.12 |
| 6.11 | | | | 9.29 | 9.25±0.01 | 0.52±0.02 | 0.72 | 14.29 | 14.22±0.01 | 1.27±0.05 | 1.91 |
| 7.01 | | | | 10.25 | 10.23±0.02 | 0.61±0.06 | 0.62 | 14.45 | | | |
| 9.88 | 9.94±0.03 | 1.41±0.06 | 2.06 | 11.26 | 11.32±0.03 | 0.68±0.07 | 0.77 | | | | |
| 11.15 | 11.16±0.07 | 1.15±0.23 | 0.54 | 12.50 | | | | **CAI ALL** | | | |
| 14.90 | | | | 14.29 | 14.24±0.16 | 4.05±0.46 | 1.18 | 9.43 | 9.30±0.01 | 0.39±0.05 | 0.16 |
| 15.43 | | | | 15.63 | | | | 9.85 | 9.76±0.01 | 0.62±0.05 | 0.53 |
| 16.20 | | | | | | | | 10.26 | 10.25±0.00 | 0.36±0.02 | 0.37 |
| 19.57 | | | | **Vigarano CAI1 INT** | **(CV3.3)** | | | 10.91 | 10.87±0.01 | 0.61±0.04 | 0.61 |
| | | | | 9.40sh | 9.29±0.04 | 0.45±0.13 | 0.10 | 11.12 | 11.22±0.01 | 0.36±0.08 | 0.16 |
| **Highly Shocked** | | | | 9.84 | 9.76±0.01 | 0.59±0.07 | 0.49 | 11.63 | 11.62±0.02 | 0.53±0.07 | 0.27 |
| 9.50sh | 9.61±0.02 | 0.94±0.03 | 0.85 | 10.25 | 10.24±0.01 | 0.27±0.02 | 0.26 | 12.37 | 12.35±0.01 | 0.91±0.03 | 1.03 |
| 10.67 | 10.40±0.01 | 0.71±0.05 | 0.67 | 10.92 | 10.92±0.03 | 0.73±0.12 | 0.58 | 14.47 | 14.66±0.01 | 0.74±0.04 | 0.55 |
| 11.25 | 11.28±0.01 | 0.97±0.03 | 1.25 | 11.21 | 11.22±0.02 | 0.14±0.10 | 0.03 | | | | |
| 13.72 | | | | 11.63 | 11.63±0.13 | 1.96±0.10 | 1.48 | | | | |
| 19.49 | 19.65±0.05 | 2.36±0.15 | 1.61 | 12.38 | 12.38±0.01 | 0.52±0.07 | 0.37 | | | | |
| | | | | 13.97 | 13.97±0.03 | 0.43±0.12 | 0.12 | | | | |
| **Maximum Shocked** | | | | 14.54 | 14.64±0.04 | 1.07±0.10 | 0.61 | | | | |
| 6.29 | | | | | | | | | | | |
| 7.01 | | | | **Vigarano CAI2 INT** | **(CV3.3)** | | | | | | |
| 10.17 | 10.19±0.01 | 0.81±0.02 | 0.64 | 9.88 | 9.87±0.01 | 0.82±0.03 | 0.81 | | | | |
| 11.27 | 11.27±0.00 | 0.80±0.01 | 1.13 | 10.25 | 10.26±0.00 | 0.16±0.17 | 0.11 | | | | |
| 11.98 | | | | 10.92 | 10.92±0.01 | 0.90±0.06 | 0.99 | | | | |
| 16.89 | | | | 11.63 | 11.70±0.02 | 0.67±0.11 | 0.34 | | | | |
| 19.49 | 19.60±0.11 | 1.78±1.05 | 0.32 | 12.38 | 11.37±0.01 | 0.70±0.03 | 0.83 | | | | |
| | | | | 13.30 | | | | | | | |
| **Average Shocked** | | | | 14.05 | 13.99±0.03 | 0.48±0.12 | 0.14 | | | | |
| 6.12 | | | | 14.62 | 14.64±0.04 | 1.07±0.14 | 0.56 | | | | |
| 6.98 | | | | | | | | | | | |
| 10.13 | 10.07±0.01 | 1.35±0.03 | 2.05 | | | | | | | | |
| 11.26 | 11.24±0.01 | 0.82±0.04 | 0.74 | | | | | | | | |
| 19.57 | 20.11±0.05 | 4.22±0.19 | 2.14 | | | | | | | | |

Table 2. Band positions of features in the laboratory spectra. In brackets: Type of chondrule or chondrite. Band: Position of peak without fitting (in µm), Barycenter: position of band after fitting a lorenzian function (in µm). FWHM: Full Width at Half Maximum, a measure of the width of the fitted feature (in µm). Int.=Intensity, strength of feature based on area of fitted band. Chon=Chondrule, INT=Interior of component, RIM= Outer part of component. CAI ALL= Average of CAI with melilite and spinel contents.

| | Type | Age (Myr) | Inner/Warm Disk | Outer/Cold Disk | Source |
|---|---|---|---|---|---|
| **Group 1** | | | | | |
| **GK Tau** | K7/M0 | 1-2 | **Am.Sil.,**Ol,Pyx | **Am.Sil.,**Ol,**Pyx**,Sil | 1,2 |
| **EX Lup** | M0 | 3 | | | 4 |
| **HD163296 (Outer)** | A1Ve | 4 | | **Am.Sil.** | 3,7,9 |
| **HD144432 (Outer)** | A9IVev | 10 | | **Am.Sil.** | 3,8,9 |
| **Group 2a** | | | | | |
| **V955 Tau** | K7 | 1-2 | **Am.Sil.,**Ol,**Pyx**,Sil | **Am.Sil.,Ol**,Pyx,Sil | 1,2 |
| **DK Tau** | K7 | 1-2 | **Am.Sil.,**Ol,Pyx,Sil | **Am.Sil.,Ol,Pyx,Sil** | 1,2 |
| **HD104237** | A4IVe+sh | 2 | **Am.Sil.,**Ol,Pyx,Sil | **Am.Sil.,**Ol,Pyx,Sil | 3,7 |
| **HD142527 (Outer)** | F7IIIe | 1 | | **Am.Sil.,Ol,Pyx** | 3,8,9 |
| **HD163296 (Inner)** | A1Ve | 4 | **Am.Sil.,Ol,Pyx** | | 3,7,9 |
| **HD144432 (Inner)** | A9IVev | 10 | **Am.Sil.,Ol,Pyx** | | 3,8,9 |
| **Group 2b** | | | | | |
| **OPH3/ RXJ1612.6-1859** | M0 | 3 | **Am.Sil.,**Ol,Pyx,Sil, | **Am.Sil.,**Ol | 6 |
| **Group 3a** | | | | | |
| **EK Cha/ RECX-5** | M3.8 | 5-9 | **Am.Sil.,Ol,Pyx,**Sil, | **Am.Sil.,**Ol,Sil, | 5 |
| **HD100546** | B9Vne | 10 | **Am.Sil.,**Ol,Pyx,Sil | **Am.Sil.**,Ol,Sil, | 3,7 |
| **HD142527 (Inner)** | F7IIIe | 1 | Am.Sil.,**Ol, Pyx** | | 3,8,9 |
| **HD144668** | A5-A7 | <1 | **Am.Sil.,Ol,Pyx,**Sil | | 8 |
| **Group 3b** | | | | | |
| **Hale Bopp** | Comet (Long period) | | **Am.Sil**.,**Ol, Pyx**,**Others** | | 10, 13, 14 |
| **9P/Tempel1** | Comet (Short Period) | | **Am.Sil.,Ol, Pyx,Others** | | 10 |
| **73P/Schwassmann-Wachmann** | Comet (Short period) | | **Am.Sil.,Ol,**Pyx,**Others** | | 11 |
| **17P/Holmes** | Comet (Short period) | | **Am.Sil.,For,Pyx,Others** | | 12 |
| **Group 4** | | | | | |
| **HD179218** | B9e | <1 | **Am.Sil.,Pyx,** Ol,Sil | **Am.Sil.,**Ol,Pyx | 3,7 |

Table 3: Information about the stars used in discussion. Types: B-G stars are Herbig type circumstellar disks with masses 2-8 times solar, K-M stars T Tauri with 0.5 to 2 times Solar. Mineralogical composition (only if quantitative data was available): Am.Sil: Amorphous silicate with olivine or pyroxene stoichiometry, Ol: crystalline olivine/forsterite; Pyx: crystalline pyroxene/enstatite; Sil: $SiO_2$.

Normal: 1-<10 wt%, **BOLD**: ≥10 wt% **BOLD UNDERLINED**:>50 wt. [1] Furlan et al.(2011) [2] Sargent et al.,2009 [3] Juhasz et al. (2010) [4] Juhasz et al. (2012) [5] Sicilia-Aguilar et al. (2009) [6] Bouwman et al. (2008) [7] van den Ancker et al. (1998) [8] van der Boekel et al. (2005) [9] van der Boekel et al. (2004) [10] Lisse et al. (2007) [11] Sitko et al. (2011) [12] Reach et al. (2010) [13] Wooden et al., 1999 [14] Wooden et al., 2000.

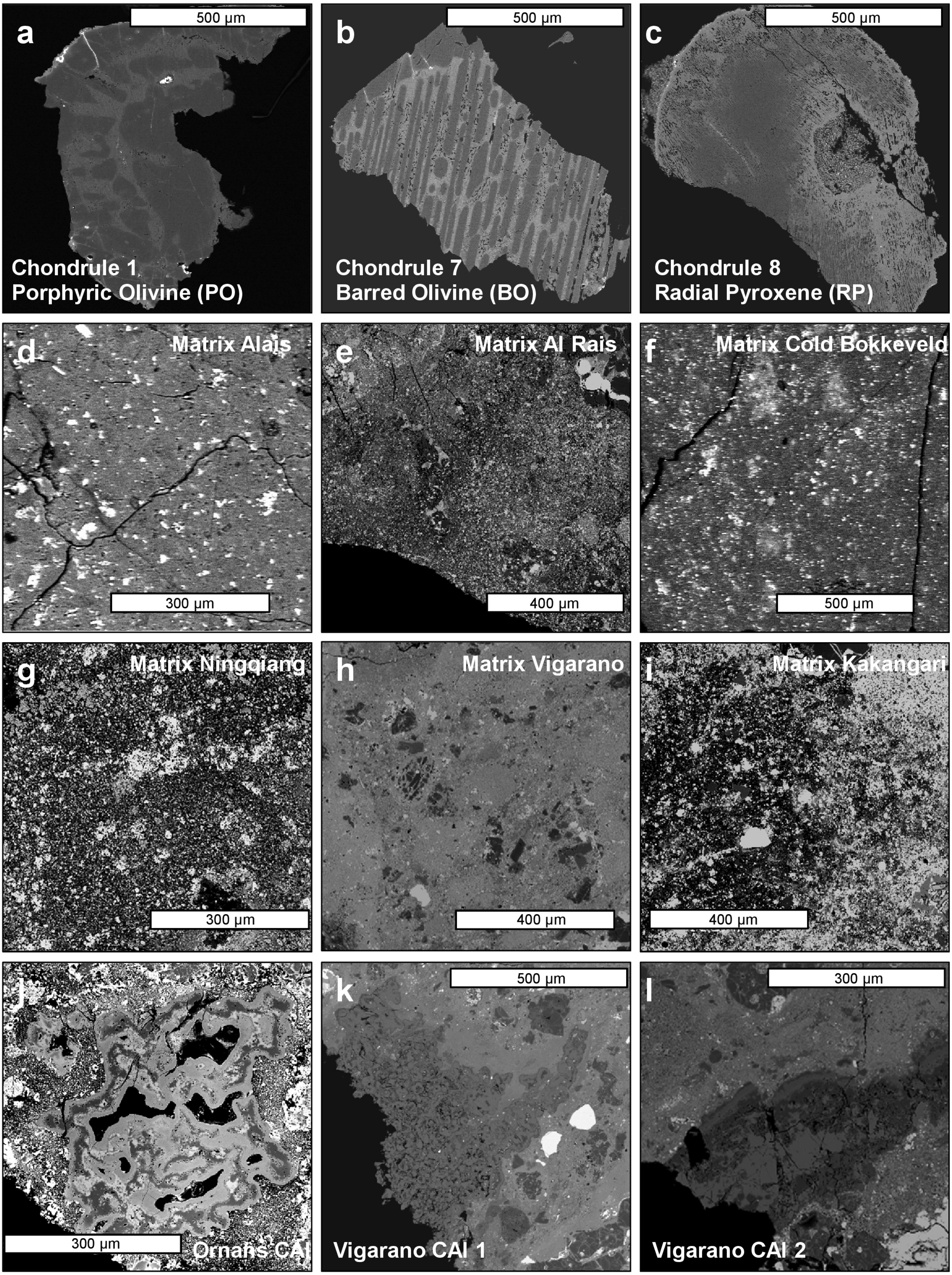

Figure 1

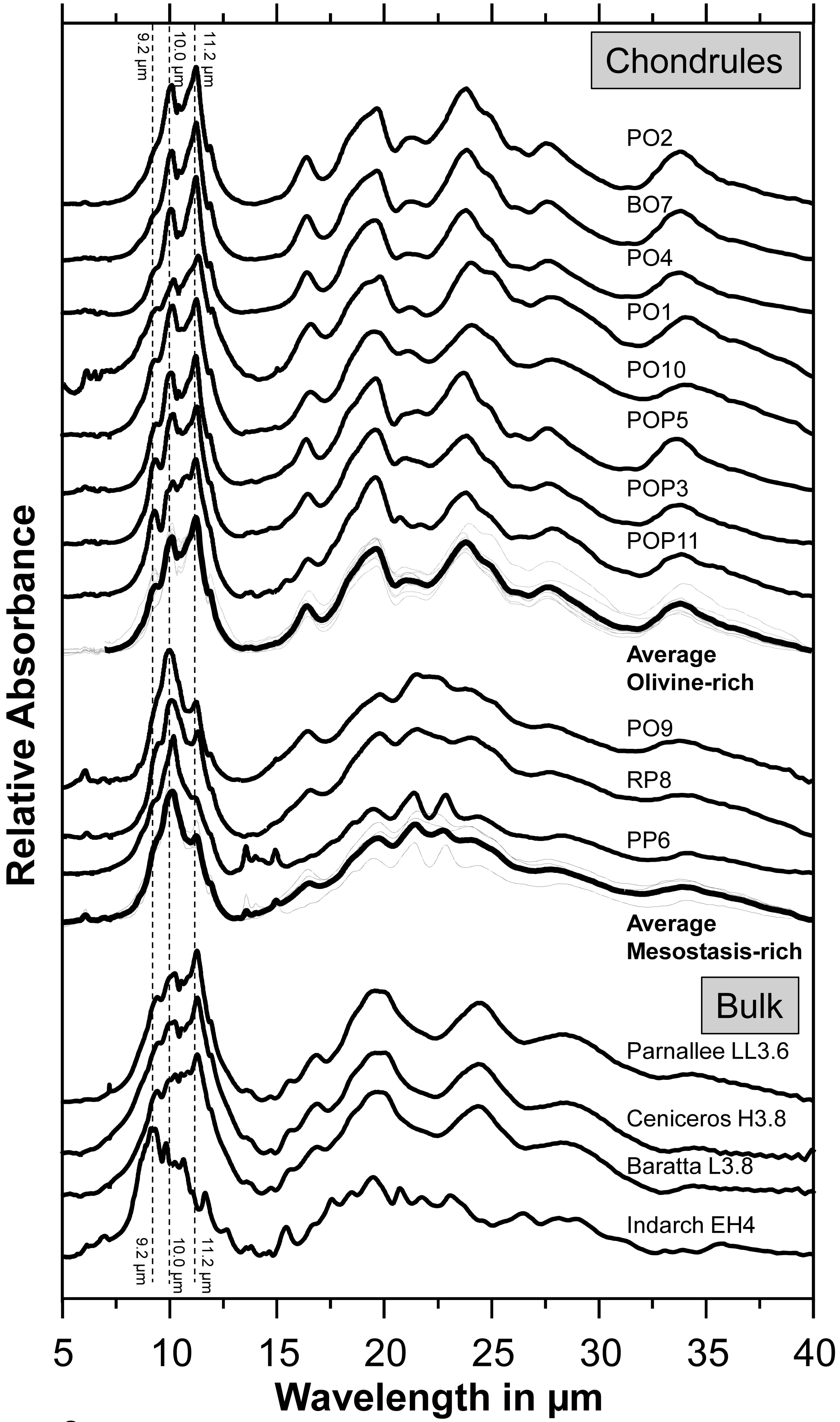

Figure 2

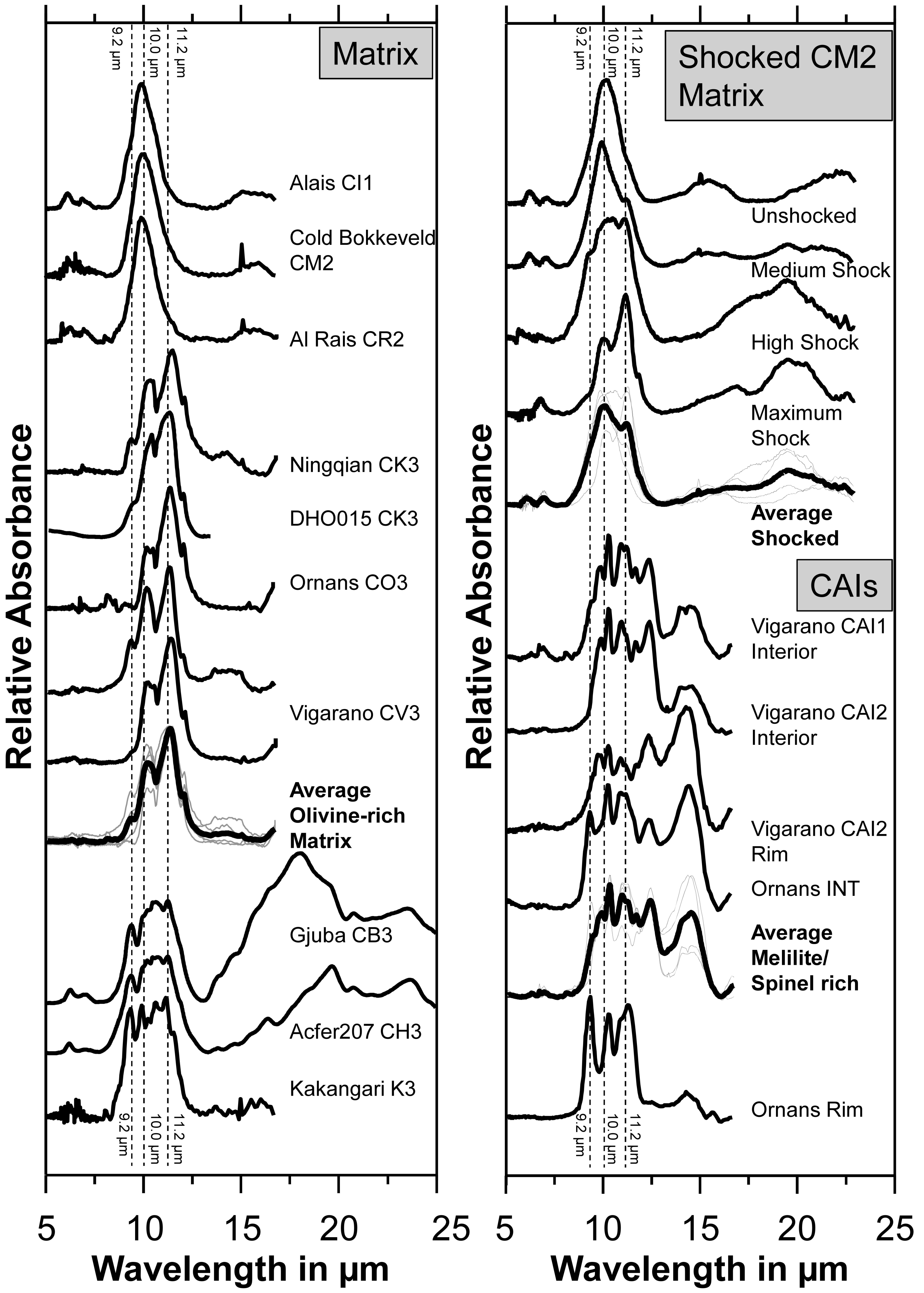

Figure 2

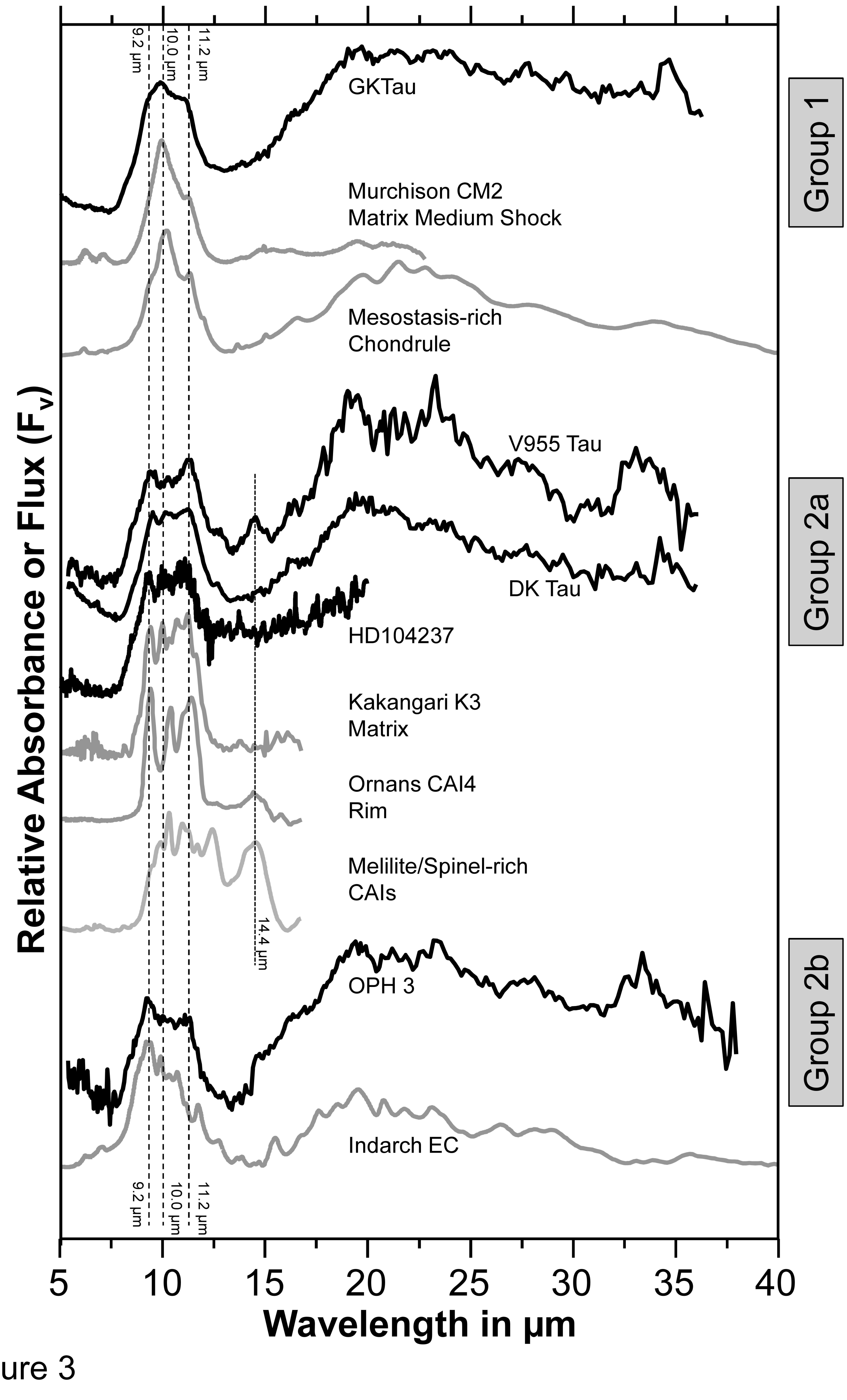

Figure 3

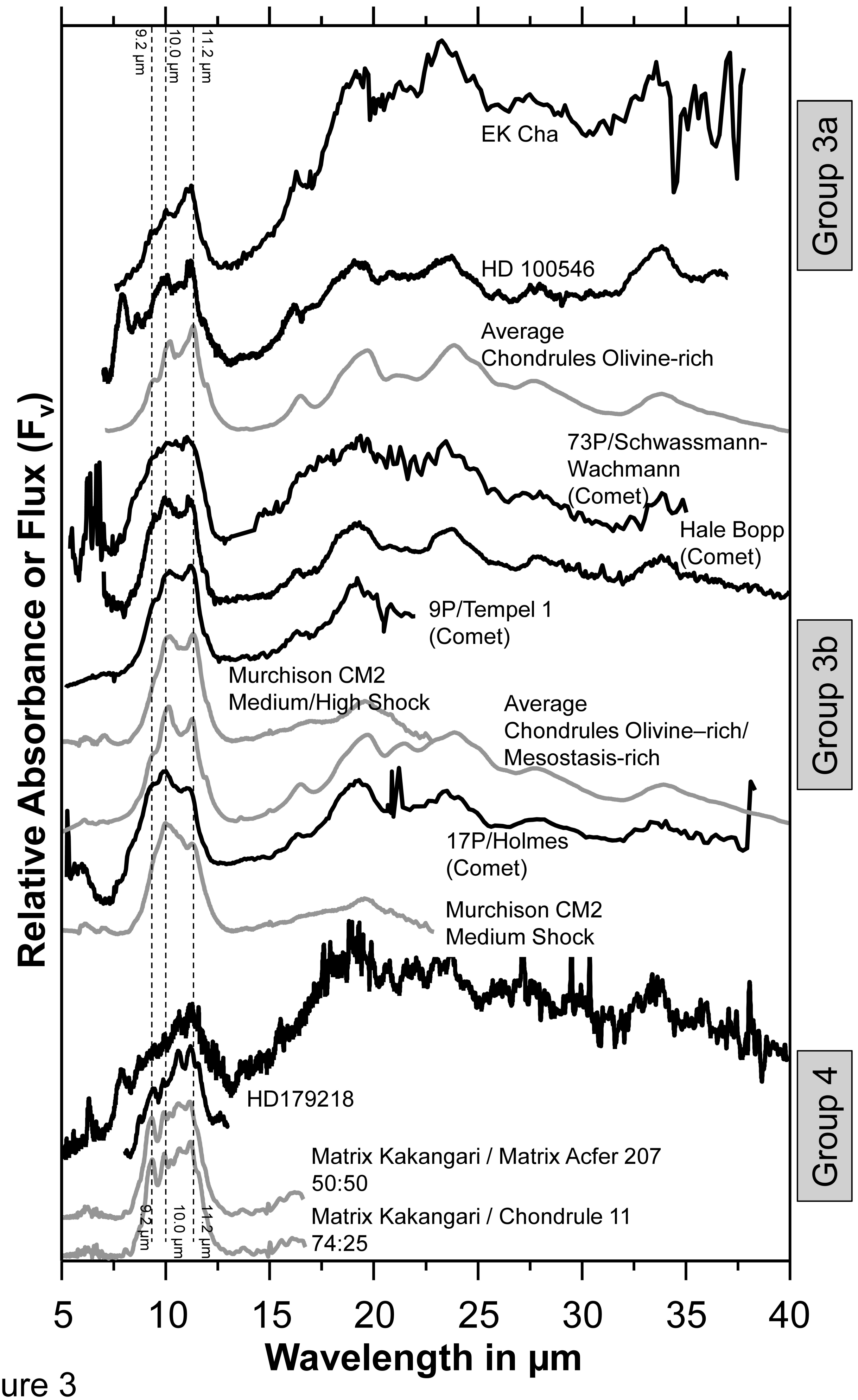

Relative Absorbance or Flux ($F_v$)
Wavelength in µm
9.2 µm
10.0 µm
11.2 µm
EK Cha
HD 100546
Average Chondrules Olivine-rich
73P/Schwassmann-Wachmann (Comet)
Hale Bopp (Comet)
9P/Tempel 1 (Comet)
Murchison CM2 Medium/High Shock
Average Chondrules Olivine–rich/ Mesostasis-rich
17P/Holmes (Comet)
Murchison CM2 Medium Shock
HD179218
Matrix Kakangari / Matrix Acfer 207 50:50
Matrix Kakangari / Chondrule 11 74:25
Group 3a
Group 3b
Group 4
5
10
15
20
25
30
35
40

Figure 3

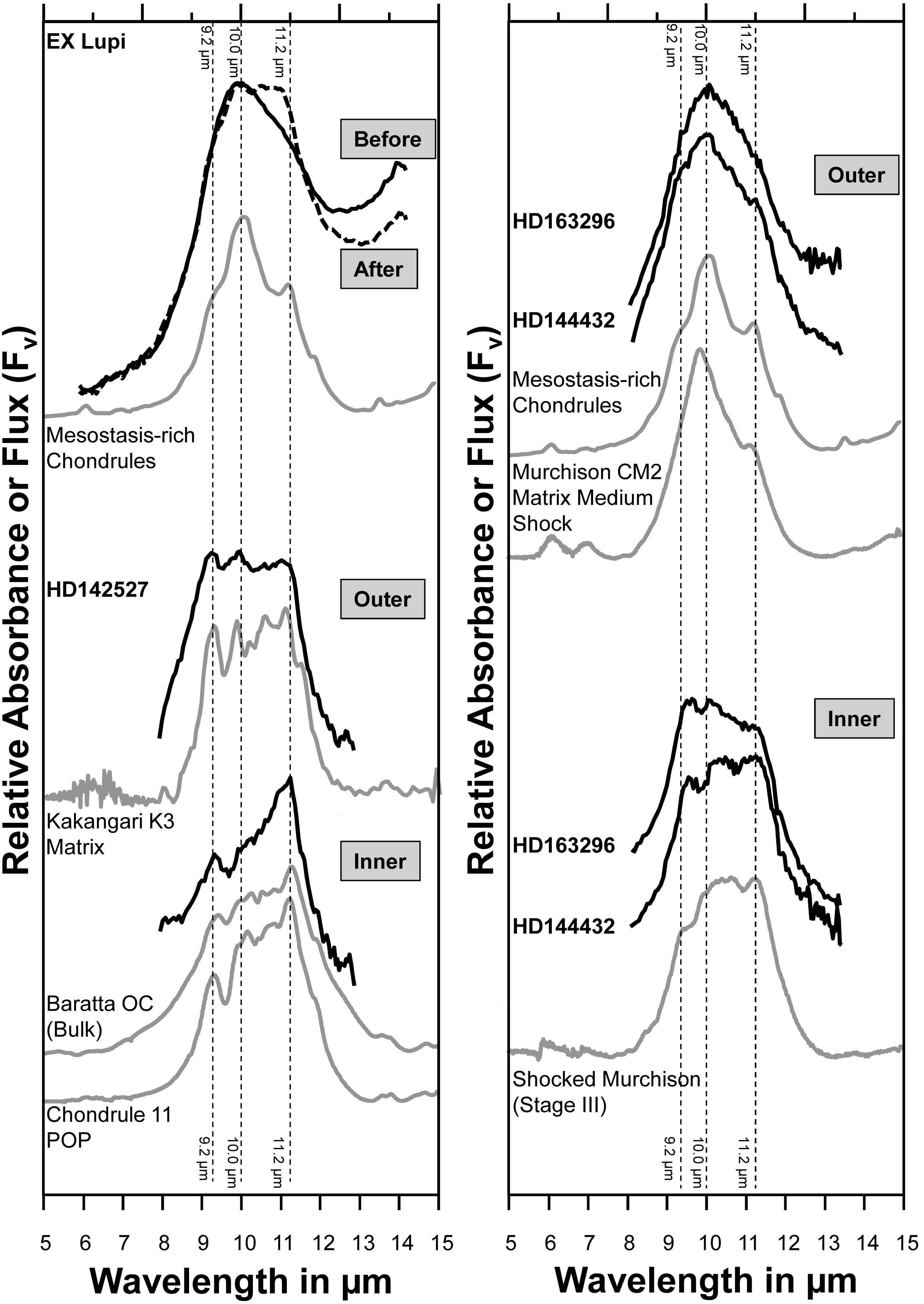

EX Lupi
9.2 µm
10.0 µm
11.2 µm
Before
After
Mesostasis-rich Chondrules
HD142527
Outer
Kakangari K3 Matrix
Inner
Baratta OC (Bulk)
Chondrule 11 POP
Outer
HD163296
HD144432
Mesostasis-rich Chondrules
Murchison CM2 Matrix Medium Shock
Inner
HD163296
HD144432
Shocked Murchison (Stage III)
Relative Absorbance or Flux ($F_v$)
Wavelength in µm
5 6 7 8 9 10 11 12 13 14 15

Figure 4

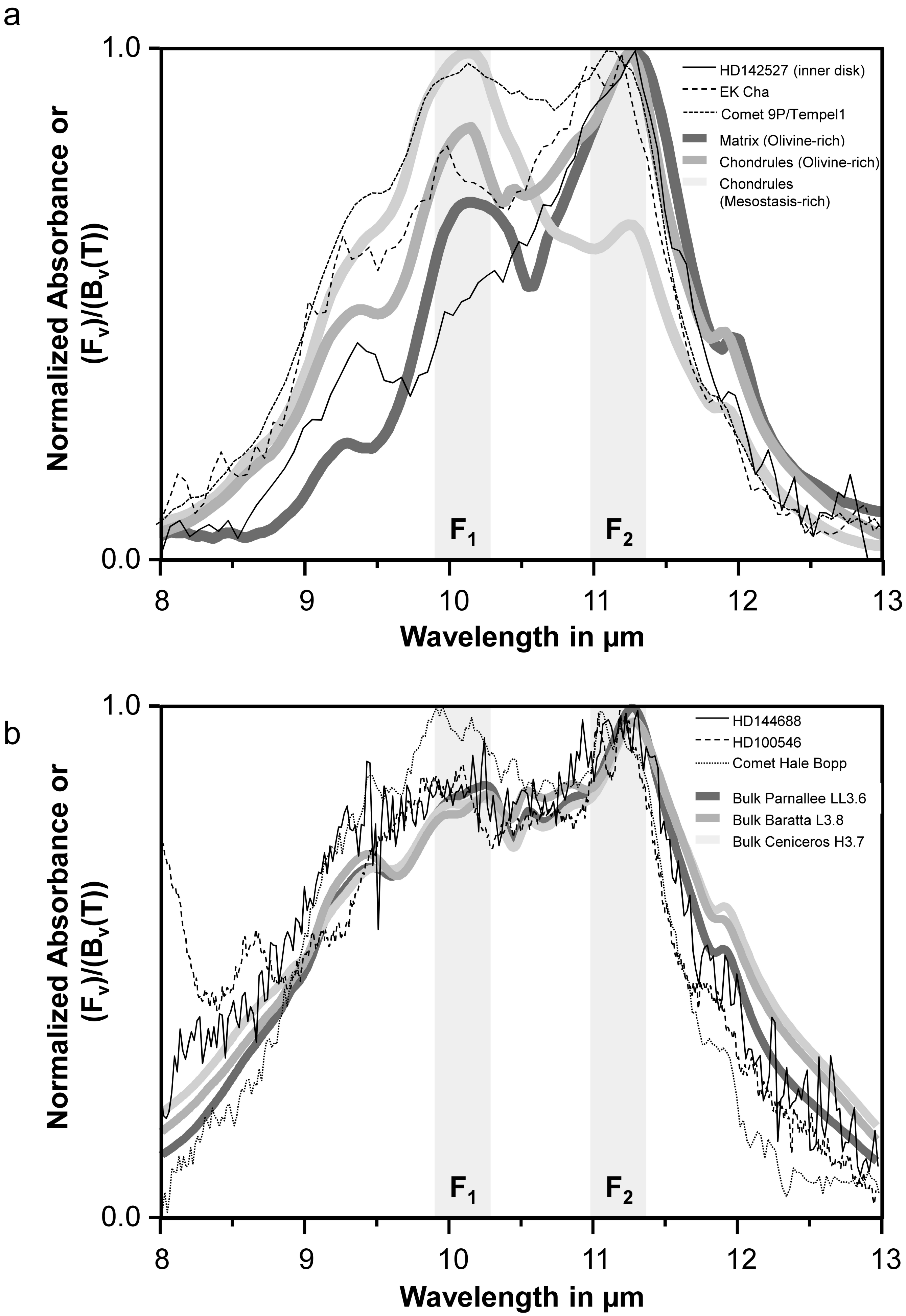

Figure 5